\documentclass[twocolumn]{aastex62}

\usepackage{amsmath}
\usepackage{graphicx}
\usepackage{lipsum}
\usepackage{hyperref}
\usepackage{mathtools, nccmath}
\usepackage{longtable}

\newcommand{\spitzer}{\textit{Spitzer}}
\newcommand{\Hii}{H\,{\sc ii}}
\newcommand{\Ciii}{C\,{\sc iii-iv} $\lambda4650$}
\newcommand{\Civ}{C\,{\sc iv} $\lambda5801\text{-}12$}
\newcommand{\Heii}{He\,{\sc ii} $\lambda4686$}
\newcommand{\Hei}{10830-\AA\ He\,{\sc i}}
\newcommand{\scol}{\mathrm{[3.6]}-\mathrm{[4.5]}}
\newcommand{\Ciiisin}{C\,{\sc iii} $\lambda4650$}
\newcommand{\Civsin}{C\,{\sc iv} $\lambda4658$}
\newcommand{\rev}{}
\newcommand{\revv}{}

\graphicspath{{./}{figures/}}

\accepted{January 4, 2021}
\submitjournal{ApJ}

\shorttitle{Extragalactic Dust-Forming WC Binaries}
\shortauthors{Lau et al.}

\begin{document}

\title{Revealing Efficient Dust-Formation at Low Metallicity in Extragalactic Carbon-Rich Wolf-Rayet Binaries}

\correspondingauthor{Ryan Lau}
\email{ryanlau@ir.isas.jaxa.jp}

\author{Ryan M. Lau}
\affil{Institute of Space \& Astronautical Science, Japan Aerospace Exploration Agency, 3-1-1 Yoshinodai, Chuo-ku, Sagamihara, Kanagawa 252-5210, Japan}
\author{Matthew J. Hankins}\affil{Department of Physical Sciences, Arkansas Tech University, 1701 N. Boulder Avenue, Russellville, AR 72801, USA}
\affiliation{Division of Physics, Mathematics, and Astronomy, California Institute of Technology, Pasadena, CA 91125, USA}
\author{Mansi M.\ Kasliwal}
\affiliation{Division of Physics, Mathematics, and Astronomy, California Institute of Technology, Pasadena, CA 91125, USA}
\author{Howard E.\ Bond}
\affiliation{Department of Astronomy \& Astrophysics, Pennsylvania State University, University Park, PA 16802, USA}
\affiliation{Space Telescope Science Institute, 3700 San Martin Drive, Baltimore, MD 21218, USA}
\author{Kishalay De}
\affiliation{Division of Physics, Mathematics, and Astronomy, California Institute of Technology, Pasadena, CA 91125, USA}
\author{Jacob E.\ Jencson}
\affiliation{Division of Physics, Mathematics, and Astronomy, California Institute of Technology, Pasadena, CA 91125, USA}
\author{Anthony F.~J.~Moffat}
\affil{Département de physique, Université de Montréal, C.P. 6128, succ. centre-ville, Montréal (Qc) H3C 3J7, Canada; Centre de Recherche en Astrophysique du Québec, Canada}
\author{Nathan Smith}
\affiliation{University of Arizona, Steward Observatory, 933 N. Cherry Avenue, Tucson, AZ 85721, USA}
\author{Peredur M. Williams}
\affil{Institute for Astronomy, University of Edinburgh, Royal Observatory, Edinburgh EH9 3HJ, UK}



\begin{abstract}

We present \textit{Spitzer}/IRAC observations of dust formation from six extragalactic carbon-rich Wolf-Rayet (WC) binary candidates in low-metallicity (Z $\lesssim0.65$ Z$_\odot$) environments using multi-epoch mid-infrared (IR) imaging data from the SPitzer InfraRed Intensive Transients Survey (SPIRITS). Optical follow-up spectroscopy of SPIRITS~16ln, 19q, 16df, 18hb, and 14apu reveals emission features from C\,{\sc iv} $\lambda5801\text{-}12$~and/or the C\,{\sc iii-iv} $\lambda4650$/He\,{\sc ii} $\lambda4686$~blend that are consistent with early-type WC stars. We identify SPIRITS~16ln as the variable mid-IR counterpart of the recently discovered colliding-wind WC4+O binary candidate, N604-WRXc, located in the sub-solar metallicity NGC~604 H\,{\sc ii}~region in M33. We interpret the mid-IR variability from SPIRITS~16ln as a dust-formation episode in an eccentric colliding-wind WC binary. SPIRITS~19q, 16df, 14apu, and 18hb exhibit absolute [3.6] magnitudes exceeding one of most IR-luminous dust-forming WC systems known, WR~104 (M$_\mathrm{[3.6]}\lesssim-12.3$). An analysis of dust formation in the mid-IR outburst from SPIRITS~19q reveals a high dust production rate of $\dot{M}_d\gtrsim2\times10^{-6}$ M$_\odot$ yr$^{-1}$, which may therefore exceed that of the most efficient dust-forming WC systems known. We demonstrate that efficient dust-formation is feasible from early-type WC binaries in the theoretical framework of colliding-wind binary dust formation if the systems host an O-type companion with a high mass-loss rate ($\dot{M}\gtrsim1.6\times10^{-6}$ M$_\odot$ yr$^{-1}$). This efficient dust-formation from early-type WC binaries highlights their potential role as significant sources of dust in low-metallicity environments.

\end{abstract}


\section{Introduction} \label{sec:intro}
A classical Wolf-Rayet (WR) star is thought to be the He-rich descendant of a massive O-type star.
WR stars are characterized by fast stellar winds ($\gtrsim1000$ km s$^{-1}$), hot temperatures ($T_*\gtrsim40000$ K), and high luminosities ($L_*\gtrsim10^5$ L$_\odot$; \citealt{Crowther2007}). 
Despite the extreme conditions around WR stars \rev{that} seem inhospitable for dust, there is a subset of carbon-rich WR (WC) stars that exhibit active dust formation \citep{Allen1972,Gehrz1974,Williams1987}. 
The earliest recorded dust-formation episode around a WC star was based on observations of infrared (IR) variability from the WC7+O5 system WR~140 (as HD 193793; \citealt{Williams1978,Hackwell1979}).
Dust formation in WC stars is facilitated by colliding-wind interactions between WC star and an OB-star companion \citep{Usov1991}. Dust formation from WC+OB binaries is therefore regulated by the orbital properties of the binary system. The link between the orbital configuration and dust formation has been directly observed in several WC+OB-star binaries. One of the most iconic WC binary dust-makers is WR~140, which is a highly eccentric system that exhibits a dust-formation episode every 7.9-yr orbital period when the stars are at their closest orbital separation \citep{Williams1990a,Usov1991,Williams2009WR140}. Dust-formation has been observed from WC binaries that exhibit orbital periods ranging from 0.7 to $\sim100$ yr \citep{Tuthill2008,Williams2019,Han2020}.

Given their capability to output dust at a high rate ($\dot{M}_d\sim10^{-10}-10^{-6}$ M$_\odot$ yr$^{-1}$; \citealt{Zubko1998,Lau2020}) and the short timescales associated with the evolutionary onset of the WR phase ($\lesssim$Myr), WC binaries present a potentially significant and early source of dust. Importantly, WC stars with a binary companion should be common based on the results from the recent spectroscopic survey by \citet{Dsilva2020}, who predict that the intrinsic multiplicity fraction of the Galactic WC population is at least $70\%$.

Recently, \citet{Lau2020} revisited the impact of dust production from WC binaries \rev{on} the interstellar medium (ISM) of galaxies in different metallicity environments representative of different epochs in cosmic time. They incorporated dust production into Binary Population and Spectral Synthesis (BPASS; \citealt{Eldridge2017}) models and compared the dust input rates from WC binaries, asymptotic-giant-branch stars (AGBs), red supergiants (RSGs), and core-collapse supernova (SNe) at different metallicities.
SNe were shown to be likely net destroyers of dust which reaffirmed results from previous studies (e.g.~\citealt{Temim2015}); however, the efficiency of dust destruction from SN shocks is still disputed (e.g.~\citealt{MartinezGonzalez2019}). \citet{Lau2020} identified the population of WC binaries as significant sources of dust with total dust production rates comparable to the population of AGB stars in solar and Large Magellanic Cloud (LMC)-like ($\sim0.5$ Z$_\odot$) metallicities with constant star formation histories. However, they acknowledged the dearth of known efficient dust-forming WC binaries in the LMC or sub-solar environments, which precludes the ability to test model predictions at sub-solar metallicities. This issue is exacerbated by the fact that most of the known dust-forming WC binaries are located within the solar circle of our Galaxy (e.g.~\citealt{Rosslowe2015,Williams2019}) indicating a bias towards studying WC systems at solar or super-solar metallicities.

There is indeed a notable trend in the spectral sub-type of WC stars at different metallicities: the cooler, late-type WC stars (WC7-9) are found in high metallicity environments, whereas the hotter, early-type WC stars (WC4-5) are found at lower metallicities \citep{Massey1998,Hadfield2007,Crowther2007}. The early-type WC stars exhibit faster wind velocities ($v_\infty\sim3000$ km s$^{-1}$) and hotter temperatures ($T_*\sim 80000$ K) than their late-type counterparts \citep{Sander2019}, which likely impacts the dust-formation efficiency from the colliding-wind mechanism \citep{Usov1991}. Notably, only a few of the 84 known dust-forming WC systems in the Galaxy host early-type WC stars according to the Galactic Wolf Rayet Calalogue V 1.25 \citep{Rosslowe2015}\footnote{\url{http://pacrowther.staff.shef.ac.uk/WRcat/}}.

\begin{figure*}[t!]
    \centerline{\includegraphics[width=0.96\linewidth]{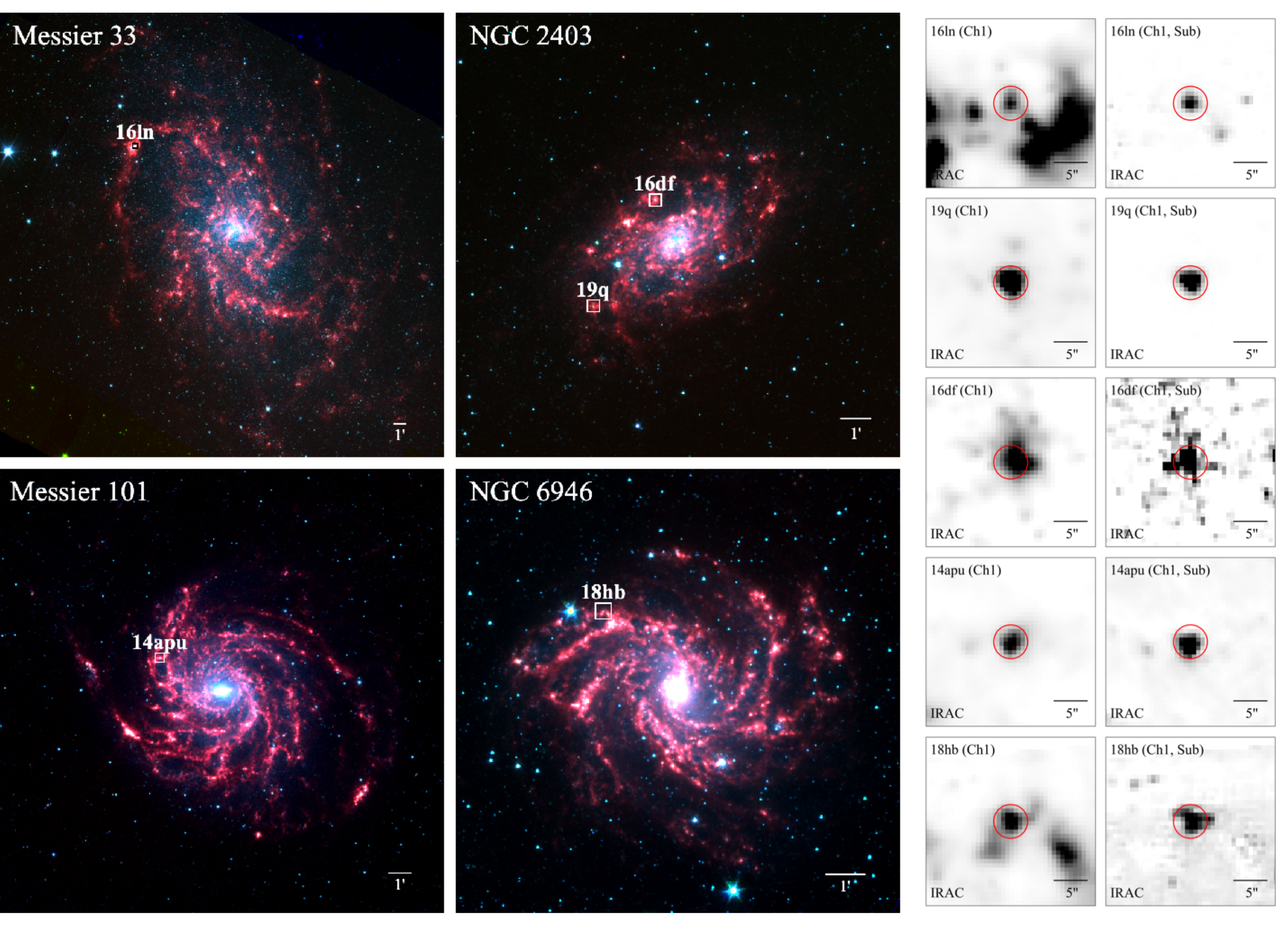}}
    \caption{(\textit{Left}) False color 3.6 (blue), 4.5 (green), and 8.0 (red) $\mu$m \spitzer/IRAC images of dust-forming WC candidate spiral galaxy hosts M33, NGC~2403, M101, and NGC~6946. (\textit{Right}) $25\times25''$ \spitzer/IRAC Channel 1 (3.6 $\mu$m) image cutout pairs of candidates SPIRITS~16ln, 19q, 16df, 14apu, and 18hb at observed IR peak and with reference image subtraction. The cutout pairs are overlaid with a $2\farcs4$-radius circle centered on the coordinates of the reference-subtracted candidate point-spread function.}
    \label{fig:WCGal}
\end{figure*}

Prior to the work presented in this paper, there were only two known dust-forming WC systems beyond our Galaxy. Both of these systems host early-type WC4 stars and are located in the Large Magellanic Cloud (LMC; \citealt{Bartzakos2001, Williams2013,Williams2019}). One of these systems, HD~36402 (= BAT99-38), exhibits correlated mid-IR and X-ray variability, which is a strong indicator of dust-formation via colliding winds \citep{Williams2013}. Spectroscopic observations of HD~36402 indicate that it is a triple system consistent with a WC4(+O?) + O8I spectral type, where the inner binary WC4+O? has a 3.03-day orbital period  \citep{Moffat1990}. Multi-epoch mid-IR observations revealed variations in dust-formation from HD~36402 with a 5.1-yr period, which is presumably associated with the outer orbit with the more-distant O8 supergiant \citep{Williams2013,Williams2019}. The only other extragalactic dust-forming WC system, HD~38030 (= BAT99-84), was recently identified from a mid-IR brightening event in 2018 \citep{Williams2019}.
The dearth of information on extragalactic and early-type dust-forming WC binaries motivates our search to discover and characterize more of these systems.

Dust formation from extragalactic WC binaries is difficult to identify due to the limitations on spatial resolution and sensitivity of mid-IR observations. \citet{Williams2019}, however, demonstrated the utility of a multi-epoch search for mid-IR variability from colliding-wind WC binaries in our Galaxy and the LMC using $\sim5$ years of archival data from the Near-Earth Object Wide-field Infrared Survey Explorer Reactivation (NEOWISE-R; \citealt{Mainzer2014}) mission. We are able to drastically expand the search for dust formation from extragalactic WC binaries by utilizing the higher spatial resolution, deeper sensitivity, and 16-yr lifetime of the \textit{Spitzer Space Telescope}.

In this paper, we present mid-IR light curves of six extragalactic dust-forming WC candidates identified from \spitzer/IRAC 3.6 and 4.5 $\mu$m imaging observations~in the SPitzer InfraRed Intensive Transients Survey (SPIRITS; \citealt{Kasliwal2017,Karambelkar2019}). One of these candidates includes the recently discovered WC+O colliding-wind binary candidate in M33, N604-WRXc \citep{Garofali2019}. For five of the six candidates, we present follow-up optical spectroscopy that reveals emission features consistent with early-type WC stars.
We perform a mid-IR color-magnitude diagram (CMD) analysis to determine if the colors exhibited by the candidates are consistent with dust formation and also compare them to other dust-forming and non dust-forming WC stars.
Lastly, we derive the dust-formation properties for the brightest mid-IR source in our sample, SPIRITS~19q, and discuss the implications and feasibility of dust formation in early-type WC binaries based on the theoretical framework of colliding-wind dust-formation presented by \citet{Usov1991}.

\begin{figure*}[t!]
    \centerline{\includegraphics[width=0.98\linewidth]{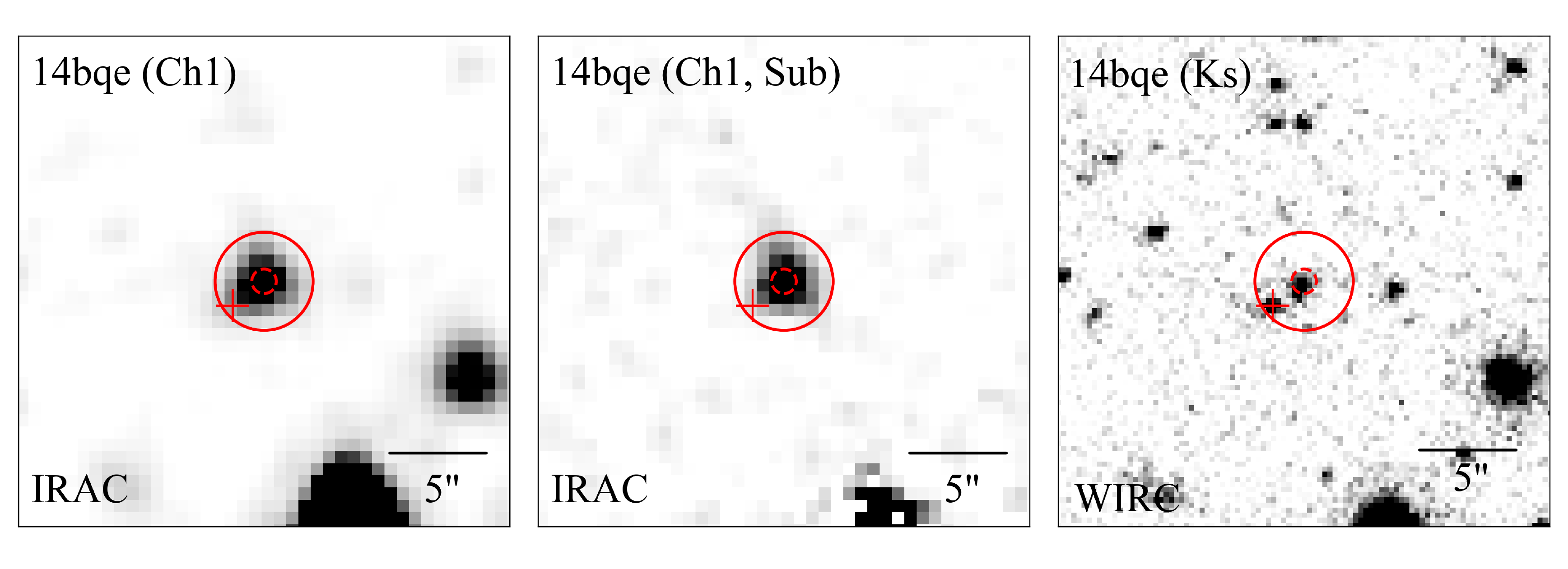}}
    \caption{ (\textit{Left}) \spitzer/IRAC Ch1 and (\textit{Center}) reference-subtracted images of SPIRITS~14bqe at IR peak. (\textit{Right}) WIRC K$_s$-band, image of SPIRITS~14bqe in IC~1613. The images cover the same $25\times25$'' field of view and are overlaid with two circles centered on SPIRITS~14bqe corresponding to the 2$\sigma$ ($0\farcs6$) positional uncertainty (dashed) and the $2\farcs4$-radius aperture used for photometry (solid). The ``+" corresponds to the position of the near-IR south-east neighbor of SPIRITS~14bqe. North is up and east is to the left in all images.}
    \label{fig:14bqe}
\end{figure*}

\section{Observations}
\label{sec:2}

\subsection{Mid-IR Photometry from SPIRITS: the SPitzer InfraRed Intensive Transients Survey}

Mid-IR variability from extragalactic dust-forming WC candidates was identified from the imaging data taken in the SPIRITS survey \citep{Kasliwal2017}. SPIRITS was a systematic, targeted survey of 194 nearby galaxies within 20 Mpc using Channels 1 (3.6 $\mu$m) and 2 (4.5 $\mu$m) of the InfraRed Array Camera (IRAC, \citealt{Fazio2004}) on the \textit{Spitzer Space Telescope} \citep{Werner2004,Gehrz2007} during its Warm Mission. SPIRITS ran from 2014 until the decommissioning of \textit{Spitzer} in Jan 2020. Observations in SPIRITS were performed with cadence baselines ranging from 1 week to 6 months and achieved a 5$\sigma$ depth of 20 mag at 3.6 $\mu$m and 19.1 mag at 4.5 $\mu$m in the Vega system. All magnitudes provided in this work are given in the Vega units.

The six dust-forming WC candidates SPIRITS~14apu, 14bqe, 16df, 16ln, 18hb, and 19q were identified from their eruptive or irregular variability and possible associations with \Hii~regions. These candidates have also been vetted through the SPIRITS image subtraction pipeline (See \citealt{Kasliwal2017}), which also included archival \textit{Spitzer} Post Basic Calibrated (PBCD) imaging data from the \textit{Spitzer Heritage Archive}\footnote{\url{https://sha.ipac.caltech.edu/applications/Spitzer/SHA/}}. 
The coordinates and host galaxies of the six SPIRITS sources are summarized in Tab.~\ref{tab:SPIRITSObs}.

\spitzer/IRAC $25''\times25''$ image cutouts and the locations of SPIRITS~16ln, 16df, 19q, 14apu, and 18hb within their host galaxies are shown in Fig.~\ref{fig:WCGal}. The \spitzer/IRAC cutouts for SPIRITS~14bqe are shown in Fig.~\ref{fig:14bqe}. SPIRITS~16ln is located in the \Hii~region NGC 604 on the outskirts of M33. SPIRITS~16df and SPIRITS~19q are both located in bright, dusty H II regions in NGC~2403. SPIRITS~14apu and SPIRITS~18hb are located within the dusty spiral arms in the outer regions of M101 and NGC~6946, respectively. Unlike the other five candidates with spiral host galaxies, SPIRITS~14bqe is located in the nearby irregular dwarf galaxy IC~1613.

Photometry was obtained using an aperture with a 4.0 PBCD mosaic pixel ($2\farcs4$) radius centered on the coordinates of the sources (Fig.~\ref{fig:WCGal} \&~\ref{fig:14bqe}).
For SPIRITS~14bqe, 16df, 19q, 14apu, and 18hb a sky background annulus extending from 4.0 to 12.0 mosaic pixels ($2\farcs4$ - $7\farcs2$) was used to subtract the diffuse galaxy background emission. Given the bright emission and crowded region around SPIRITS~16ln, a 24.0 to 40.0 mosaic pixel ($40\farcs0$ - $66\farcs7$) sky annulus was used to subtract the background emission. The Ch1 and Ch2 aperture correction factors for SPIRITS~16ln photometry were 1.208 and 1.220, respectively. For the other five candidates with the 4.0 to 12.0 mosaic pixel background sky annuli, aperture correction factors of 1.215 and 1.233 were adopted for the Ch1 and Ch2 photometry, respectively. These aperture correction factors are consistent with the values provided in the \textit{Spitzer}/IRAC manual for the adopted aperture radii and background annuli. \textit{Spitzer}/IRAC Ch1 and Ch2 photometry and observation dates of all six dust-forming WC candidates are provided in Tab.~\ref{tab:LCTab}.

\begin{deluxetable*}{llllll}
\tablecaption{SPIRITS Dust-Forming WC Candidates}
\tablewidth{0.98\linewidth}
\tablehead{SPIRITS ID & J2000 Coordinates & Galaxy & Distance ($\mu)$ & Distance Ref.& Follow-up Obs. }
\startdata
14bqe & 01:05:01.95	+02:08:51.3 & IC 1613 & 724 kpc ($24.30$) & Ha17 & WIRC (Imaging) \\ 
16ln & 01:34:32.61 +30:47:04.3 & Messier 33 & 840 kpc ($24.63$) & Gi13 & LRS2 \\ 
19q & 07:37:18.21 +65:33:49.0 & NGC 2403 & 3.2 Mpc ($27.51$) & Ra11 & LRIS, NIRES, LRS2 \\ 
16df & 07:36:58.08 +65:37:23.9 & NGC 2403 & 3.2 Mpc ($27.51$) & Ra11 & LRIS \\
14apu & 14:03:31.77 +54:22:25.1 & Messier 101 & 6.4 Mpc ($29.04$) & Sh11 & LRIS \\
18hb & 20:35:07.68 +60:11:16.2 & NGC 6946 & 7.72 Mpc ($29.44$) & An18 & LRS2 \\
\enddata
\tablecomments{Coordinates, galaxy hosts, galaxy host distances, and follow-up observations conducted of the six SPIRITS dust-forming WC candidates. Abbreviations correspond to the following references: Ha17 - \citet{Hatt2017}, Gi13 - \citet{Gieren2013}, Ra11 - \citet{RS2011}, Sh11 - \citet{Shappee2011}, An18 - \citet{Anand2018}. }
\label{tab:SPIRITSObs}
\end{deluxetable*}

\subsection{Optical Follow-up Spectroscopy}
\subsubsection{Keck I/LRIS}

Optical follow-up spectroscopy of SPIRITS~19q, 16df, and 14apu were obtained using the Low Resolution Imaging Spectrometer (LRIS; \citealt{Goodrich2003}) on the Keck I telescope. LRIS observations of SPIRITS~19q were conducted on 2019 Apr 3, and the observations of SPIRITS~16df and 14apu were both taken on 2020 Mar 23. All three observations used the D560 dichroic with the standard $1\farcs0\times175''$ longslit slitmask and the 400/3400 grism on the blue side and the 400/8500 grating on the red side. The resulting spectral resolving power on the blue ($\lambda=3500-6000$ \AA) and red ($\lambda=5500-10300$ \AA) sides were $R\sim600$ and $R\sim1000$, respectively. 

\begin{figure}[t!]
    \centerline{\includegraphics[width=0.98\linewidth]{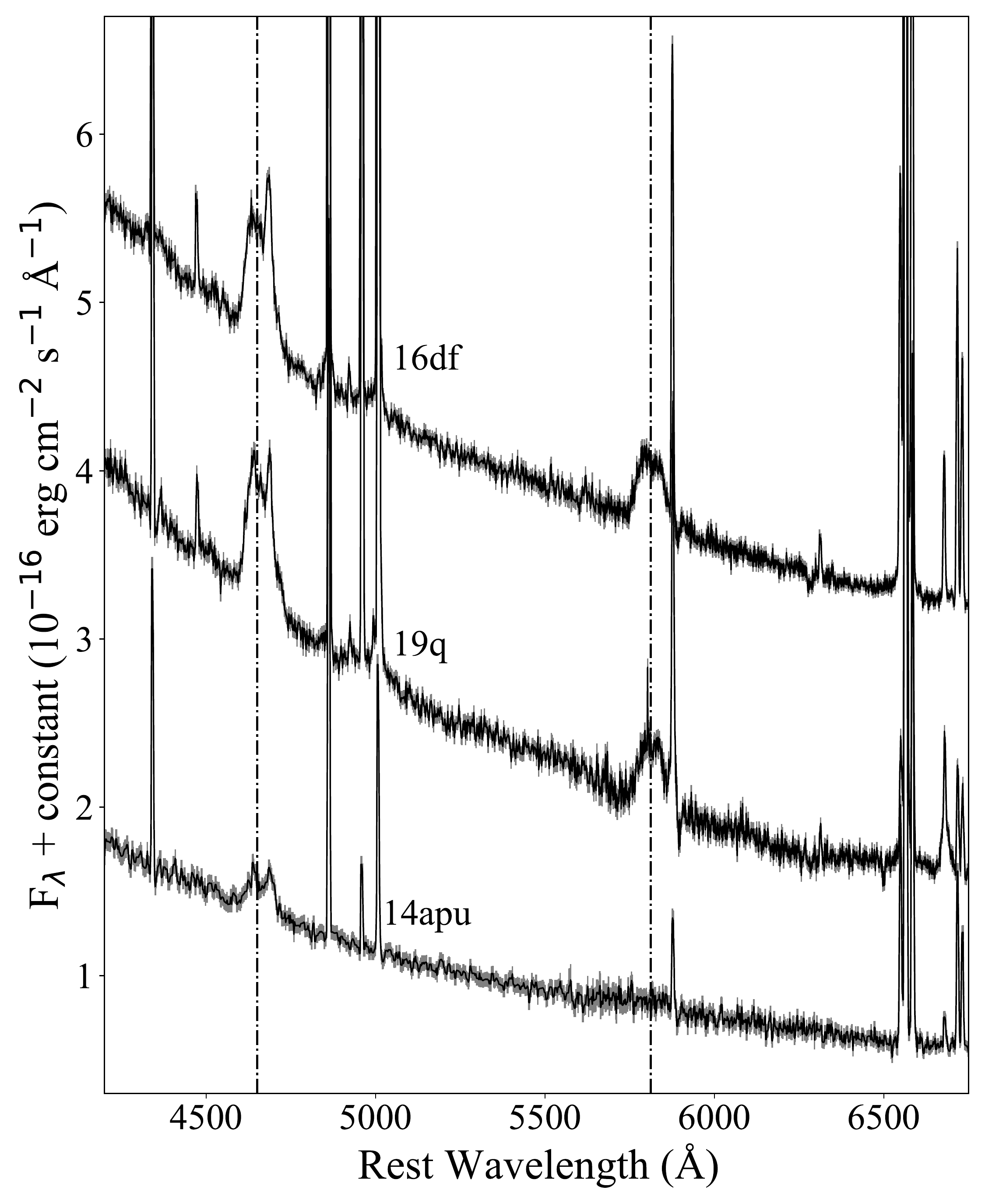}}
    \caption{Reduced and flux calibrated Keck I/LRIS spectra of SPIRITS~16df, 19q, and 14apu overlaid with dashed lines corresponding to the wavelengths of \Civ~and the \Ciii/\Heii~blend. Narrow emission features are likely associated with the underlying or nearby \Hii~regions. The spectrum of SPIRITS~16df is shown with a continuum offset of $2\times10^{-16}$ erg cm$^{-1}$ s$^{-1}$ \AA$^{-1}$.}
    \label{fig:LRIS}
\end{figure}

The total exposure times averaged between the red and blue sides for SPIRITS~19q, 16df, and 14apu were 580 s, 750 s, and 300 s respectively. Data were reduced using the fully automated Keck/LRIS reduction software reduction package ``LPipe" \citep{Perley2019}. The flux calibration for SPIRITS~19q was performed using the standard star G191-B2B, and the flux calibrations for SPIRITS~16df and SPIRITS~14apu were performed using Feige34. The reduced and flux-calibrated LRIS spectra of SPIRITS~19q, 16df, and 14apu are shown in Fig.~\ref{fig:LRIS}. The spectra reveal broad features corresponding to the \Civ~and/or the \Ciiisin/\Civsin/\Heii~blend as well as a series of narrow emission lines likely originating from underlying/nearby \Hii~regions. Hereafter, we refer to the combined \Ciiisin/\Civsin~components as \Ciii.

\subsubsection{Hobby-Eberly Telescope/LRS2}
Optical follow-up spectroscopy was also conducted with the Low-Resolution Spectrograph 2 (LRS2; \citealt{Chonis2016}), an integral field spectrograph with a field of view of $6''\times12''$ on the 10-m Hobby-Eberly Telescope (HET). Observations of SPIRITS~16ln, 18hb and 19q were obtained with the blue spectrograph of LRS2, LRS2-B, which covers $3700-4700$ \AA~($R\approx 1900$) and $4600-7000$ \AA~($R\approx 1100$) in its ``UV Arm'' and ``Orange Arm,'' respectively. The observations centered on SPIRITS~16ln, 18hb and 19q were taken on 2019 Sept 8, 2019 Aug 20, and 2019 Oct 29, respectively. The total exposure times for SPIRITS~16ln, 18hb and 19q were 750 s, $3\times1400$ s, and $2\times750$ s, respectively.

\begin{figure}[t!]
    \centerline{\includegraphics[width=0.98\linewidth]{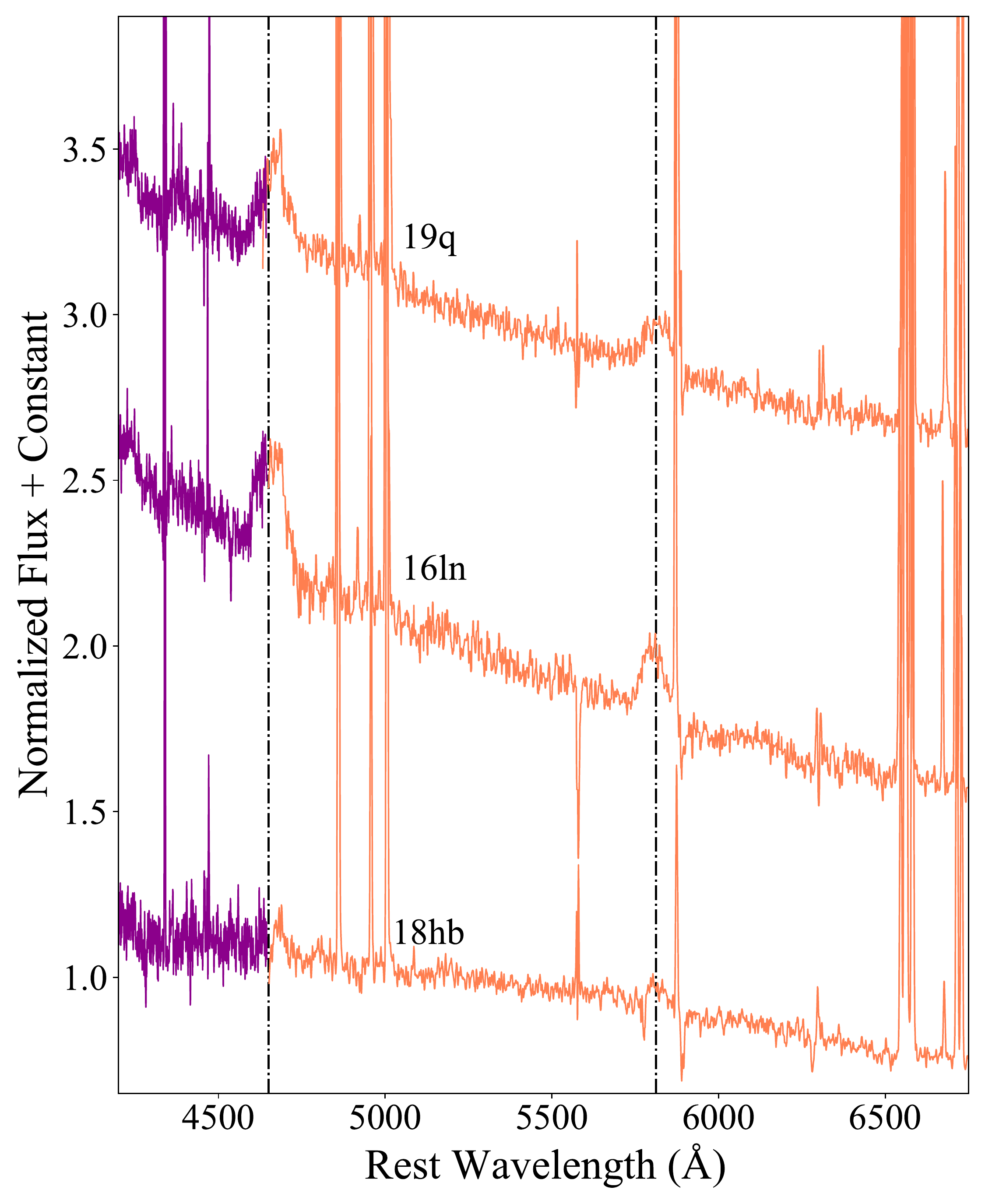}}
    \caption{Reduced HET/LRS2 spectra of SPIRITS~19q, 16ln, and 18hb showing the flux normalized to the peak emission of the \Ciii/\Heii~blend and offset for clarity. The overlaid with dashed lines corresponding to the wavelengths of \Civ~and the \Ciii/\Heii~blend. Narrow emission features are likely associated with the underlying or nearby \Hii~regions.}
    \label{fig:HET}
\end{figure}

The spectra from the UV and Orange channels were reduced separately using the LRS2 pipeline ``Panacea\footnote{\url{https://github.com/grzeimann/Panacea}}." Since observations were obtained in non-photometric conditions, the analysis of the LRS2 spectra of SPIRITS~16ln, 18hb, and 19q focuses on the relative flux rather than the absolute flux. The spectra were normalized to the peak emission of the \Ciii/\Heii~emission complex in the Orange channel. 
Due to normalization offsets between the UV and Orange channels, the UV channel emission is offset to match the slope of the continuum emission in the Orange channel. The reduced and normalized LRS2 spectra of SPIRITS~16ln, 18hb, and 19q are shown in Fig.~\ref{fig:HET}. Similar to the Keck I/LRIS spectra, the LRS2 spectra show broad features corresponding to the \Civ~and/or the \Ciii/\Heii~blend as well as a series of narrow emission lines that likely originate from underlying/nearby \Hii~regions.

\begin{deluxetable*}{lllccllll}
\tablecaption{\textit{Spitzer}/IRAC Photometry of SPIRITS Dust-Forming WC Candidates}
\tablewidth{0pt}
\tablehead{MJD & F$_\mathrm{3.6}$ (mJy) & F$_\mathrm{4.5}$ (mJy) & [3.6] & [4.5] & [3.6] - [4.5] & M$_\mathrm{[3.6]}$ & M$_\mathrm{[4.5]}$ & PI (Prog ID)}
\startdata
\textbf{SPIRITS16ln}\\ 
53001.39 & 0.632 $\pm$ 0.030 & 0.600 $\pm$ 0.026 & 14.12 $\pm$ 0.05 & 13.69 $\pm$ 0.05 & 0.43 $\pm$ 0.07 & -10.50 & -10.93 & Houck (63)\\
53013.72 & 0.615 $\pm$ 0.029 & 0.593 $\pm$ 0.026 & 14.15 $\pm$ 0.05 & 13.70 $\pm$ 0.05 & 0.45 $\pm$ 0.07 & -10.47 & -10.92 & Gehrz (5)\\
53208.96 & 0.646 $\pm$ 0.030 & 0.624 $\pm$ 0.026 & 14.10 $\pm$ 0.05 & 13.65 $\pm$ 0.05 & 0.45 $\pm$ 0.07 & -10.52 & -10.97 & Gehrz (5)\\
53233.18 & 0.646 $\pm$ 0.030 & 0.603 $\pm$ 0.026 & 14.10 $\pm$ 0.05 & 13.69 $\pm$ 0.05 & 0.41 $\pm$ 0.07 & -10.52 & -10.93 & Gehrz (5)\\
53391.68 & 0.618 $\pm$ 0.029 & 0.597 $\pm$ 0.026 & 14.14 $\pm$ 0.05 & 13.70 $\pm$ 0.05 & 0.45 $\pm$ 0.07 & -10.48 & -10.92 & Gehrz (5)\\
53607.18 & 0.645 $\pm$ 0.030 & 0.595 $\pm$ 0.026 & 14.10 $\pm$ 0.05 & 13.70 $\pm$ 0.05 & 0.40 $\pm$ 0.07 & -10.52 & -10.92 & Gehrz (5)\\
53770.66 & 0.621 $\pm$ 0.030 & 0.573 $\pm$ 0.026 & 14.14 $\pm$ 0.05 & 13.74 $\pm$ 0.05 & 0.40 $\pm$ 0.07 & -10.48 & -10.88 & Gehrz (5)\\
57111.70 & 0.973 $\pm$ 0.029 & 0.979 $\pm$ 0.026 & 13.65 $\pm$ 0.03 & 13.16 $\pm$ 0.03 & 0.49 $\pm$ 0.04 & -10.97 & -11.46 & Kasliwal (11063)\\
57118.84 & 0.969 $\pm$ 0.029 & 0.971 $\pm$ 0.025 & 13.66 $\pm$ 0.03 & 13.17 $\pm$ 0.03 & 0.49 $\pm$ 0.04 & -10.96 & -11.45 & Kasliwal (11063)\\
57139.32 & 0.952 $\pm$ 0.028 & 0.942 $\pm$ 0.025 & 13.68 $\pm$ 0.03 & 13.20 $\pm$ 0.03 & 0.47 $\pm$ 0.04 & -10.94 & -11.42 & Kasliwal (11063)\\
57314.62 & 0.798 $\pm$ 0.029 & 0.777 $\pm$ 0.025 & 13.87 $\pm$ 0.04 & 13.41 $\pm$ 0.04 & 0.46 $\pm$ 0.05 & -10.75 & -11.21 & Kasliwal (11063)\\
57322.09 & 0.809 $\pm$ 0.029 & 0.780 $\pm$ 0.025 & 13.85 $\pm$ 0.04 & 13.41 $\pm$ 0.04 & 0.45 $\pm$ 0.05 & -10.77 & -11.21 & Kasliwal (11063)\\
57336.48 & 0.780 $\pm$ 0.029 & 0.770 $\pm$ 0.025 & 13.89 $\pm$ 0.04 & 13.42 $\pm$ 0.04 & 0.47 $\pm$ 0.05 & -10.73 & -11.20 & Kasliwal (11063)\\
57491.30 & 0.711 $\pm$ 0.029 & 0.682 $\pm$ 0.026 & 13.99 $\pm$ 0.04 & 13.55 $\pm$ 0.04 & 0.44 $\pm$ 0.06 & -10.63 & -11.07 & Kasliwal (11063)\\\hline
\textbf{SPIRITS19q}\\
53285.07 & 0.353 $\pm$ 0.009 & 0.324 $\pm$ 0.008 & 14.75 $\pm$ 0.03 & 14.36 $\pm$ 0.03 & 0.39 $\pm$ 0.04 & -12.76 & -13.15 & Van Dyk (226)\\
... & ... & ... & ... & ... & ... & ... & ... & ...\\
\enddata
\tablecomments{\spitzer/IRAC 3.6 and 4.5 $\mu$m photometry (in Jy and magnitudes), $\scol$ color, and absolute magnitudes with 1$\sigma$ uncertainties of the six dust-forming WC candidates SPIRITS~16ln, 19q, 16df, 14apu, 18hb, and 14bqe. The PI names and program ID numbers of the \spitzer/IRAC programs associated with the observations are also provided. All magnitudes are given in the Vega system. A full version of this table is shown in Tab.~\ref{tab:LCTabFull}.}
\label{tab:LCTab}
\end{deluxetable*}

\subsection{Near-IR Spectroscopy with Keck II/NIRES}

\begin{figure*}[t!]
    \centerline{\includegraphics[width=0.98\linewidth]{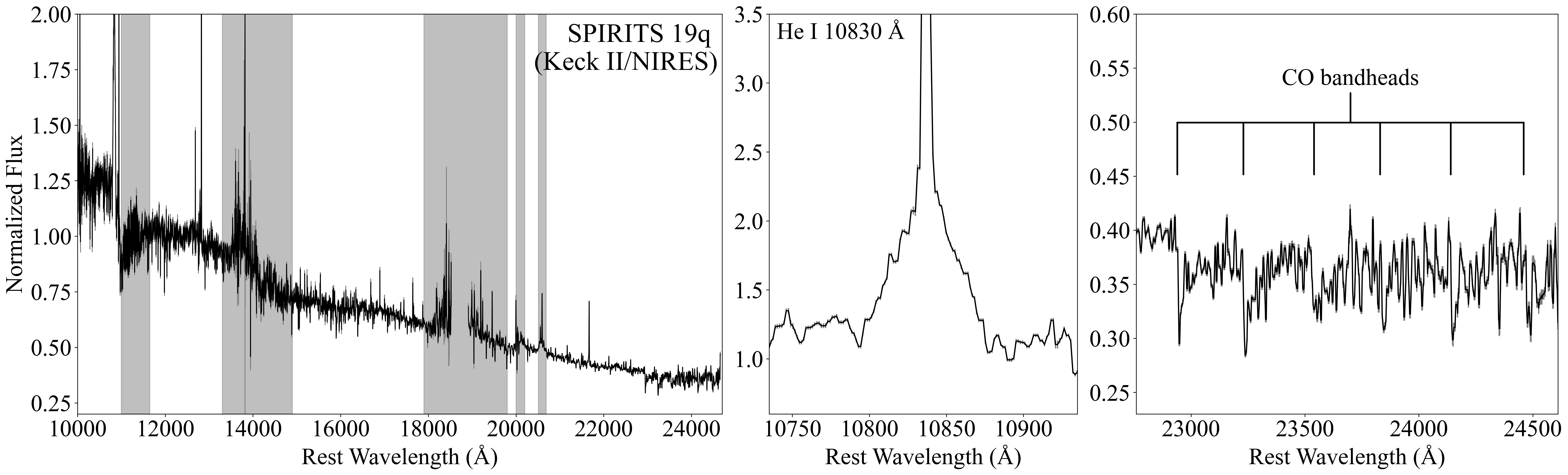}}
    \caption{Reduced Keck II/NIRES spectrum of SPIRITS~19q normalized to the J-band continuum emission and overlaid with telluric absorption bands (in gray). Zoomed panels of the normalized spectrum on the \Hei~feature and CO bandheads are shown to the right.}
    \label{fig:NIRES}
\end{figure*}

We obtained a near-IR spectrum of SPIRITS\,19q with the Near-Infrared Echellette Spectrometer\footnote{\url{https://www2.keck.hawaii.edu/inst/nires/}} (NIRES) on the Keck II Telescope on 2019 March 13. NIRES uses a $0\farcs55$ slit and
provides wavelength coverage from 9500 to 24600~\AA\ across five spectral orders with a mean spectral resolving power of $R = 2700$.
During the observations, we obtained eight individual 300 s exposures for a total integration time of 2400 s. 
The target was nodded along the slit between exposures in a standard ABBA pattern to allow for accurate subtraction of the
sky background. Observations of the A0V telluric standard star HIP~32549 near the target position were also taken immediately preceding the science target observation for flux calibration and correction of the strong near-IR telluric absorption features. The data were reduced, including flat-fielding, wavelength calibration, background subtraction, and 1D spectral extractions steps, using a version of the IDL-based data reduction package Spextool developed by \citet{Cushing2004}, updated by M.\ Cushing specifically for NIRES. Telluric corrections and flux calibrations were performed with the standard-star observations using the method developed by \citet{Vacca2003} implemented with the IDL tools \textsc{xtellcor} or \textsc{xtellcor\_general} developed by \citet{Cushing2004} as part of Spextool. The reduced spectrum normalized to the J-band (1.1 - 1.3 $\mu$m) continuum flux is shown in Fig.~\ref{fig:NIRES} with zoomed panels on the broadened \Hei~line and molecular CO bands in absorption. 

\subsection{Near-IR Imaging with P200/WIRC}
Near-IR imaging of SPIRITS~14bqe in IC~1613 were taken as part of the ground-based near-IR follow-up campaign for SPIRITS using the Wide Field Infrared Camera (WIRC; \citealt{Wilson2003}) on the 200-inch telescope at Palomar Observatory (P200). Images of IC~1613 were obtained in the J- and Ks-band filters on WIRC on 2016 Oct 11. The data were sky subtracted and then calibrated using 2MASS stars in the field of view. The measured J and Ks magnitude (in Vega) of SPIRITS~14bqe was $J=18.94\pm0.04$ and $K_s= 17.90\pm0.06$. The Ks-band image of SPIRITS~14bqe is shown in Fig.~\ref{fig:14bqe} overlaid with its $2\sigma$ ($0\farcs6$) positional uncertainty and the $2\farcs4$ aperture used for the \spitzer/IRAC photometry.

\section{Results and Analysis}
\label{sec:3}

\subsection{Identifying Dust-Formation from \spitzer/IRAC~Mid-IR Colors}
\label{sec:Color}

Mid-IR photometry is valuable for distinguishing hot dust ($T_\mathrm{d}\sim800$ K) from the free-free emission by the ionized stellar winds of WC stars \citep{Williams2019}. Here, we demonstrate how the \spitzer/IRAC Ch1 and Ch2 color can be used to identify an IR excess around WC stars consistent with thermal emission from circumstellar dust.

The \spitzer~Ch1 and Ch2 color ($\scol$) corresponding to measured Ch1 and Ch2 fluxes of $\mathrm{F}_{\nu,\mathrm{Ch1}}$ and $\mathrm{F}_{\nu,\mathrm{Ch2}}$ can be expressed as

\begin{equation}
\scol=2.5\,\mathrm{Log}\left(\frac{\mathrm{F}_{\nu,\mathrm{Ch2}}}{\mathrm{F}_{\nu,\mathrm{Ch1}}}\right)+2.5\,\mathrm{Log}\left(\frac{\mathrm{F}_{\nu0,\mathrm{Ch1}}}{\mathrm{F}_{\nu0,\mathrm{Ch2}}}\right),
\label{eq:col}
\end{equation}
\noindent
where $\mathrm{F}_{\nu0,\mathrm{Ch1}}$ and $\mathrm{F}_{\nu0,\mathrm{Ch2}}$ are the zero magnitude fluxes at Ch1 and Ch2, respectively\footnote{$\mathrm{F}_{\nu0,\mathrm{Ch1}}=280.9$ Jy and $\mathrm{F}_{\nu0,\mathrm{Ch1}}=179.7$ Jy}.
Free-free emission from ionized dense winds dominates the IR emission from dust-free WC stars and can be approximated as a power-law: $F^{ff}_\nu\propto\nu^{0.96}$ \citep{Morris1993}. 
Based on this free-free emission power-law, the \spitzer~$\scol$ color from a dust-free WC star wind can be approximated as
\begin{equation}
(\scol)_{ff}\approx0.24.
\end{equation}

\noindent This is consistent with the observed mid-IR colors of dust-free WC stars in the LMC \citep{Bonanos2009,Williams2019}.

Optically thin, mid-IR thermal dust emission at a temperature $T_\mathrm{d}$ can be expressed as a modified blackbody with a power-law: $F^{\mathrm{d}}_\nu\propto B_\nu(T_d)\,\nu^{\beta}$, where $\beta$ is the power-law emissivity index and $B_\nu$ is the Planck function. Since the circumstellar dust around WC stars is believed to be carbon-rich \citep{Cherchneff2000}, an emissivity index consistent with amorphous carbon dust is adopted: $\beta=1.2$ \citep{Zubko1996}. Note that dust-forming WC stars do not exhibit prominent mid-IR CO emission \citep{vdh1996} and thus do not expect any contamination from these features in the observed [4.5] photometry. From Eq.~\ref{eq:col}, the predicted \spitzer~$\scol$ for thermal dust emission at a temperature $T_\mathrm{d}\lesssim800$ K can be approximated as 

\begin{equation}
(\scol)_{\mathrm{d}}\gtrsim0.57,
\label{eq:IRdust}
\end{equation}
\noindent
which is redder than the dust-free WC stars dominated by free-free emission.
We therefore use Eq.~\ref{eq:IRdust} as a criterion for identifying thermal emission from circumstellar dust.

Due to the spatial resolution of \spitzer/IRAC ($\mathrm{FWHM}\sim1\farcs7$) and the location of the dust-forming WC candidates in crowded and/or dusty regions, the mid-IR photometry is likely contaminated by emission from nearby stars or nebulosity. 
The absolute mid-IR magnitudes from the aperture photometry of the WC candidates (Tab.~\ref{tab:LCTabFull}) indeed exceed the absolute magnitudes of individual WC stars in the Large Magellanic Cloud measured by \spitzer/IRAC in the Surveying the Agents of the Galaxy's Evolution (SAGE) program \citep{Meixner2006,Bonanos2009}.
In order to remove the contaminating emission from the WC candidates with excessive mid-IR brightness, we utilize the multi-epoch \spitzer~coverage to characterize and subtract the ``quiescent", non-variable emission. 
Table~\ref{tab:WC} provides the nearby \Hii~or star-forming regions and summarizes the properties of the dust-forming WC candidates.

\subsection{Episodic Dust-Formation from the WC Binary SPIRITS16ln/N604-WRXc in M33} 
\label{sec:16ln}

\begin{figure}[t!]
    \centerline{\includegraphics[width=0.98\linewidth]{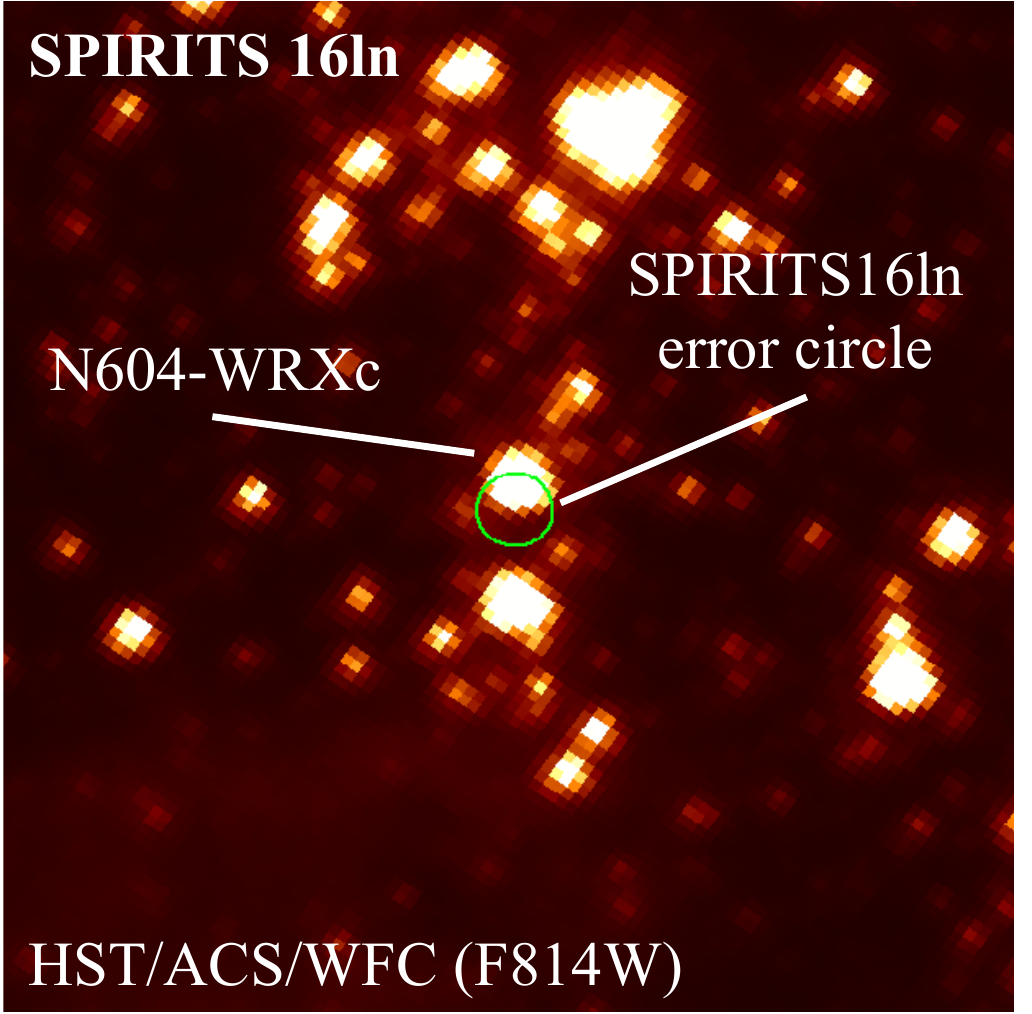}}
    \caption{HST/ACS/WFC F814W image of SPIRITS~16ln/N604-WRXc taken on 2017 Aug 4 \citep{Garofali2019} overlaid with the $1\sigma$ ($0\farcs165$) positional error circle of SPIRITS~16ln. The length and width of the image cutout is $2\farcs4\times2\farcs4$. North is up and east is to the left.  }
    \label{fig:16ln}
\end{figure}

SPIRITS~16ln is located in the massive \Hii~region NGC 604 within the nearby galaxy M33 at a distance of 840 kpc ($\mu=24.63$; \citealt{Gieren2013}). NGC 604 exhibits a sub-solar metallicity Z $=0.65$ Z$_\odot$ \citep{Vilchez1988,Margini2007,Garofali2019} comparable to that of the LMC. The \textit{Spitzer}-derived position of SPIRITS~16ln is consistent within $0\farcs13$ of the WC4+O candidate colliding-wind binary system N604-WRXc \citep{Garofali2019}. 
The early-type WC emission features shown in the HET/LRS2 spectra of SPIRITS~16ln (Fig.~\ref{fig:HET}) are notably consistent with the spectra of N604-WRXc presented by \citep{Garofali2019}.
In Fig.~\ref{fig:16ln}, we show the $1\sigma$ positional error circle of SPIRITS~16ln after an astrometric alignment between the \spitzer~image and the \textit{Hubble Space Telescope} (HST) ACS/WFC image of N604-WRXc taken with the F814W filter on 2017 Aug 4 \citep{Garofali2019}.
The position of SPIRITS~16ln is indeed consistent with N604-WRXc, and we therefore claim that SPIRITS~16ln is associated with N604-WRXc.

\begin{figure*}[t!]
    \centerline{\includegraphics[width=0.98\linewidth]{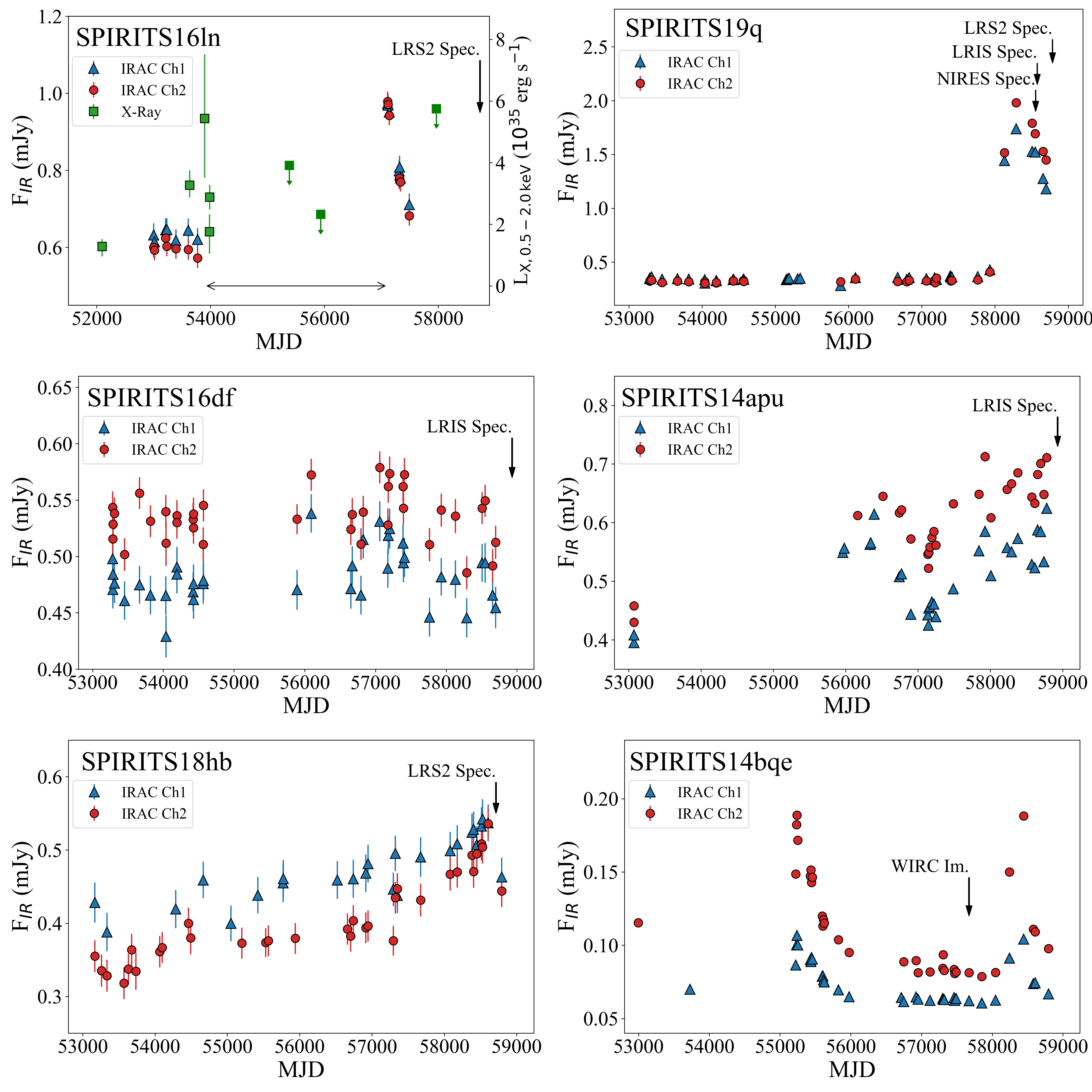}}
    \caption{\spitzer/IRAC light curves at Channel 1 (3.6 $\mu$m) and 2 (4.5 $\mu$m) of the six dust-forming WC candidates identified in SPIRITS. The \spitzer~light curve of SPIRITS~16ln/N604-WRXc is overlaid with the unabsorbed X-ray luminosity (0.5 - 2.0 keV) measurements and upper limits from \textit{Chandra} and \textit{XMM-Newton} of N604-WRXc by \citet{Garofali2019}; the horizontal double-sided arrow corresponds to the length of time ($\sim8.8$ yr) between the X-ray and mid-IR emission peaks. Dates of follow-up observations with LRIS, LRS2, and/or WIRC are indicated on the light curves.}
    \label{fig:IRLC}
\end{figure*}

SPIRITS~16ln was observed by \textit{Spitzer} in quiescence between 2003 Dec 28 - 2006 Feb 4 (MJD 53001 - 53770). Between 2015 Mar 30 - 2016 Apr 13 (MJD 57111 - 57491), SPIRITS~16ln was observed to be fading from a mid-IR outburst that should have occurred between 2006 Feb and 2015 Mar. Figure~\ref{fig:IRLC} shows the mid-IR light curve of SPIRITS~16ln overplotted with the unabsorbed X-ray fluxes of N604-WRXc reported by \citet{Garofali2019}. During quiescence the median 3.6 and 4.5 $\mu$m emission was F$_{3.6}=0.632\pm0.030$ mJy (M$_\mathrm{[3.6]}=-10.50\pm0.05$) and F$_{4.5}=0.597\pm0.026$ mJy (M$_\mathrm{[4.5]}=-10.92\pm0.05$), respectively. 
Unfortunately, the X-ray flux peak from N604-WRXc was observed in 2006 Jun 9, $\sim4$ months after the final \textit{Spitzer} observation of SPIRITS~16ln in quiescence. It is therefore unclear if the mid-IR and X-ray emission peaks are correlated.

The peak mid-IR emission from SPIRITS~16ln was observed on 2015 Mar 30 (MJD 57111), where the measured fluxes were F$_{3.6}=0.973\pm0.029$ mJy (M$_\mathrm{[3.6]}=-10.97\pm0.03$) and F$_{4.5}=0.979\pm0.026$ mJy (M$_\mathrm{[4.5]}=-11.46\pm0.03$). 
Given the rapid fading following the peaks in both the mid-IR and X-ray emission from SPIRITS~16ln/N604-WRXc, it is difficult to constrain a possible recurrence period of the outbursts. However, a lower limit of $\gtrsim2.2$ yr can be estimated from the time-span of the \spitzer~observations where SPIRITS~16ln was in quiescence.  Semi-contemporaneous mid-IR and X-ray flux peaks are associated with periastron passage in dust-forming colliding-wind binaries (e.g.~\citealt{Williams1990a}). Under the assumption that the mid-IR and X-ray peaks are indeed correlated, the upper limit of the outburst period is $\lesssim8.8$ yr (Fig.~\ref{fig:IRLC}).

\begin{figure*}[t!]
    \centerline{\includegraphics[width=0.9\linewidth]{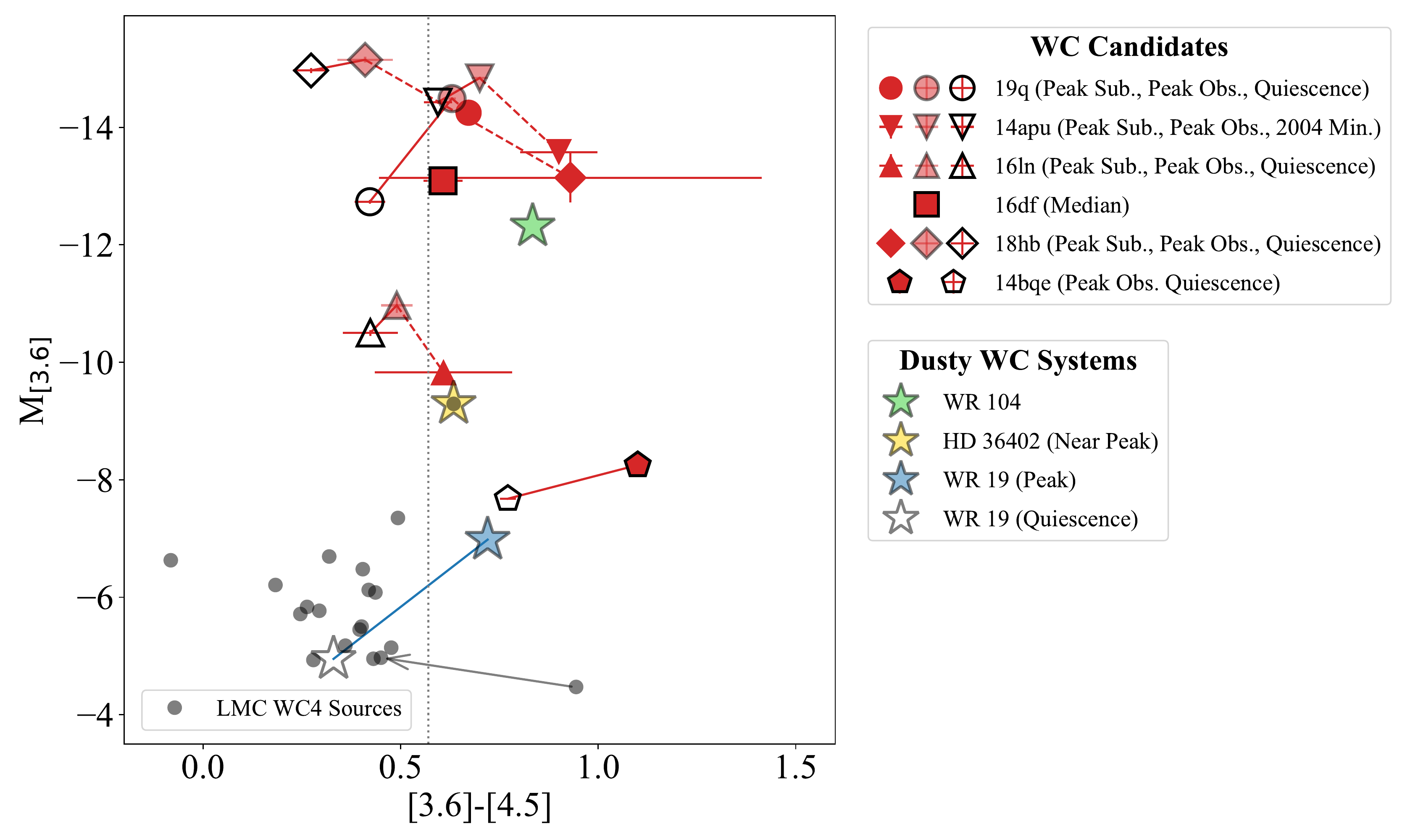}}
    \caption{Mid-IR color-magnitude diagram of the six dust-forming WC candidates, LMC WC4 sources \citep{Bonanos2009}, and three known dusty WC systems: WR~104, HD~36402 near its IR peak, and WR~19 at its IR peak. Sources to the right of the vertical dotted line exhibit colors consistent with thermal dust emission at a temperature $T_\mathrm{d}\lesssim800$ K ($\scol\gtrsim0.57$, Eq.~\ref{eq:col}). Unfilled markers correspond to source colors and magnitudes at quiescence or IR minimum. For \rev{SPIRITS 19q, 14apu, 16ln, 18hb, 14bqe} and WR~19, the filled markers \rev{with borders} correspond to \rev{observed magnitudes and colors} at their IR peaks, \rev{whereas} border-less markers indicate where the quiescent \rev{and/or underlying background} emission has been subtracted. The values of $\scol$ and M$_\mathrm{[3.6]}$ for WR~104 were derived from IR spectroscopy taken by the Short Wavelength Spectrometer (SWS) on the Infrared Space Observatory (ISO) \citep{vdh1996}. IR peak and quiescent mid-IR colors and absolute magnitudes of WR~19 were obtained from $L'$- and $M$-band images taken by the IRAC1 camera on the ESO 2.2m telescope in 1998 March \citep{Veen1998} and from NEOWISE-R W1 and W2 photometry \citep{Williams2019}, respectively. Distances toward WR~19 and WR~104 were adopted from \citet{Rate2020}. The revised \spitzer~$\scol$ and M$_\mathrm{[3.6]}$ of the apparent red LMC WC4 outlier, HD~32125, is indicated by the arrow (See Sec.~\ref{Sec:Outlier}).}
    \label{fig:CMD}
\end{figure*}

The \spitzer/IRAC~$\scol$ color and absolute [3.6] magnitude of SPIRITS~16ln during quiescence and at its emission peak is shown in the mid-IR CMD in Fig.~\ref{fig:CMD}. Since the quiescent emission contributes a significant fraction of the observed mid-IR flux at the outburst peak, Fig.~\ref{fig:CMD} also shows the color and magnitude at the outburst peak where the median emission from the quiescent phase has been subtracted.
The subtracted absolute [3.6] magnitude at peak emission is M$_\mathrm{[3.6]}=-9.8$ and the mid-IR color is $\scol \sim0.6$, which is consistent with hot dust emission (Eq.~\ref{eq:IRdust}).
The red-ward shift in the mid-IR colors during outburst is also indicative of dust formation. 
The subtracted $\scol$ color and absolute [3.6] magnitude exhibited by SPIRITS~16ln during its observed peak is also consistent with the variable dust-forming WC4 system HD~36402 around its mid-IR peak \citep{Bonanos2009,Williams2013}.

Based on our analysis and the previously reported X-ray variability \citep{Garofali2019}, we claim that SPIRITS~16ln/N604-WRXc is an episodic dust-forming colliding-wind binary hosting a WC4 star. We attribute the rapid decrease in the mid-IR emission during outburst \rev{to the fading and cooling of dust as it disperses after a brief formation period during periastron passage in a highly 
eccentric binary orbit.} 
This process is demonstrated in several Galactic dust-forming colliding-wind systems such as WR~19 and WR~140 \citep{Williams2009WR140,Williams2009WR19}. The interpretation of a binary with a high orbital eccentricity in SPIRITS~16ln is also supported by the X-ray variability reported by \citet{Garofali2019} since the X-ray luminosity in colliding-wind binaries is highly dependent on orbital separation (e.g.~\citealt{Stevens1992}). SPIRITS~16ln/N604-WRXc is therefore the first extragalactic dust-forming colliding-wind WC binary identified beyond the LMC.

\subsection{Candidate Dust-Forming WC Stars in NGC~2403: SPIRITS~19q \& SPIRITS~16df}

\subsubsection{Host \Hii~Region Environments}

SPIRITS~19q and 16df are located in dusty and bright \Hii~regions within the spiral galaxy NGC~2403 at a distance of 3.2 Mpc ($\mu=27.51$; \citealt{RS2011}). Their locations within NGC~2403 are shown in Fig.~\ref{fig:WCGal} and their \spitzer~mid-IR light curves are presented in Fig.~\ref{fig:IRLC}. SPIRITS~19q and 16df are coincident with the VS~51 and VS~41 \Hii~regions \citep{Veron1965}, respectively. VS~51 is located at a de-projected distance of 3.5 kpc from the nuclear core of NGC~2403 and VS~41 is located slightly closer to the core at a de-projected distance of 2.6 kpc \citep{Garnett1997}. 

VS~51 and VS~41 exhibit oxygen abundances of 12 + log(O/H)$_\mathrm{VS51}$ = $8.43\pm0.13$ and 12 + log(O/H)$_\mathrm{VS41}$ = $8.49\pm0.04$\footnote{These are the oxygen abundances derived by \citet{Moustakas2010} from the \citet{Pilyugin2005} calibration, which is applicable to high excitation star-forming regions such as VS~51 and VS~41.} \citep{Moustakas2010}, both of which are less than the solar oxygen abundance of 12 + log(O/H)$_\odot$ = $8.69\pm0.05$ \citep{Asplund2009}. This implies that VS~51 and VS~41 host sub-solar metallicity environments where $\mathrm{Z}_\mathrm{VS51}\sim 0.6$ Z$_\odot$ and $\mathrm{Z}_\mathrm{VS41}\sim 0.5$ Z$_\odot$, which are similar to the LMC and NGC~604. Both regions also exhibit high 0.5 - 2.0 keV X-ray luminosities of L$_\mathrm{X}>10^{35}$ erg s$^{-1}$ that are most likely powered by supernovae and/or stellar winds from massive stars \citep{Yukita2010}. 
Optical spectroscopy of VS~41 (as NGC~2403-V) obtained by \citet{Drissen1999} reveal WR features consistent with emission from the \Civ~and the \Ciii/\Heii~blend. 
\citet{Drissen1999} notably identify six WR stars in VS~41, four of which are confined to the $\sim1\farcs5$ around the bright central core of the \Hii~region that also contains a compact star cluster.

\begin{figure*}[t!]
    \centerline{\includegraphics[width=0.90\linewidth]{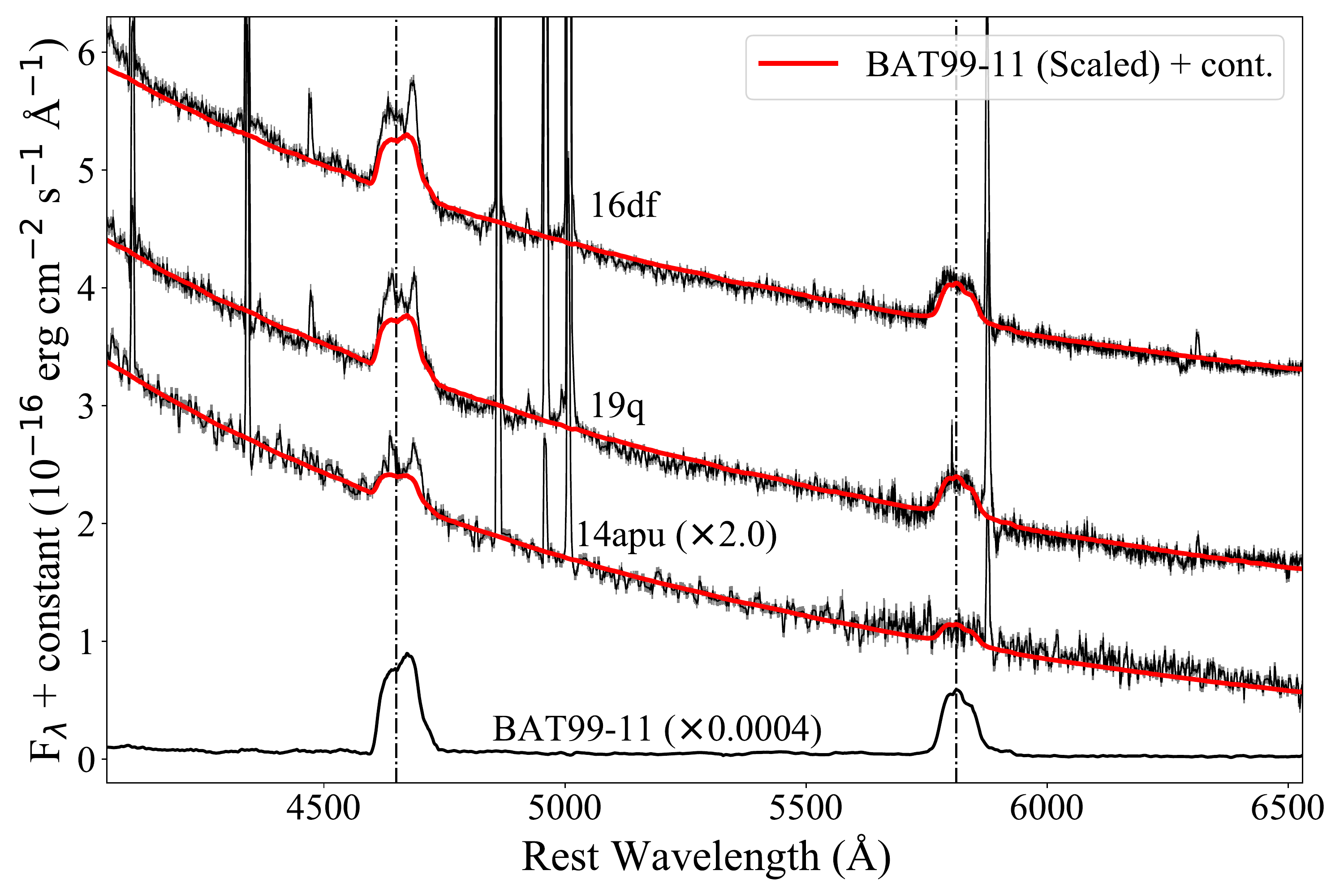}}
    \caption{Keck I/LRIS spectra of SPIRITS~16df, 19q, and 14apu with the wavelength of the \Civ~and \Ciii/\Heii~features indicated by the dot-dashed line. The spectrum of SPIRITS~16df is shown with a continuum offset of $2\times10^{-16}$ erg cm$^{-1}$ s$^{-1}$ \AA$^{-1}$, and the spectrum of SPIRITS~14apu is scaled up by a factor of two and is shown with a continuum offset of $-0.6\times10^{-16}$ erg cm$^{-1}$ s$^{-1}$ \AA$^{-1}$. Each candidate is overlaid with the spectrum of the dust-free LMC WC4 star BAT99-11 scaled to the distance of the candidate host galaxies and combined with a power-law fit of the continuum emission of the LRIS spectra.}
    \label{fig:LRISMod}
\end{figure*}

\subsubsection{SPIRITS~19q}

Rising mid-IR emission from SPIRITS~19q was captured by \spitzer~on 2017 Jun 26 (MJD 57930) which peaked almost exactly a year later on 2018 Jun 20 (MJD 58289) where it exhibited absolute mid-IR magnitudes of M$_\mathrm{[3.6]}=-14.49\pm0.01$ and M$_\mathrm{[4.5]}=-15.12\pm0.01$ (Fig.~\ref{fig:IRLC}, Tab.~\ref{tab:LCTabFull}). The final \spitzer~observations of SPIRITS~19q were obtained on 2019 Aug 2 (MJD 58697) and show that its mid-IR emission has faded at a rate of $\sim0.4$ mag yr$^{-1}$ at 3.6 $\mu$m and $\sim0.3$ mag yr$^{-1}$ at 4.5 $\mu$m after the peak. 

 SPIRITS~19q was in a quiescent phase for at least $\sim12$ years between 2004 Oct 7 - 2017 Jan 10 (MJD 53285 - 57764) before the 2018 outburst. During quiescence, the median mid-IR emission from SPIRITS~19q was F$_{3.6}=0.345\pm0.009$ mJy and F$_{4.5}=0.326\pm0.008$ mJy (M$_{[3.6]}=-12.73\pm0.03$ and M$_{[4.5]}=-13.16\pm0.03$). The bright mid-IR absolute magnitudes during quiescence (M$_{[3.6]}<-12.5$ and M$_{[4.5]}=-13.0$) suggest that SPIRITS~19q is located within an \Hii~region or star cluster (e.g.~\citealt{Lau2019}), which is also supported by \spitzer~imaging observations that reveal slightly extended emission from 19q (Fig.~\ref{fig:WCGal}).

Figure~\ref{fig:CMD} shows the absolute [3.6] magnitude and $\scol$ color of SPIRITS~19q during quiescence and at its mid-IR peak with and without the quiescent emission subtracted. SPIRITS~19q demonstrates a clear red-ward shift in its $\scol$ color from quiescence to its outburst phase, which suggests dust formation is associated with the outburst. The quiescence-subtracted mid-IR color at its emission peak is $\scol=0.67\pm0.01$, consistent with thermal dust emission as well as the mid-IR colors exhibited by SPIRITS~16ln during its outburst and HD~36402. 
Figure~\ref{fig:CMD} also shows the \rev{observed} mid-IR color and magnitude of WR~104, a well-studied persistent dust-forming WC9+OB binary in our Galaxy that is one of the most IR-luminous dust-forming WC systems known \citep{Tuthill1999,Soulain2018,Lau2020}.
Interestingly, the quiescence-subtracted absolute [3.6] magnitude of SPIRITS~19q during outburst, M$_{[3.6]}=-14.25\pm0.01$, exceeds WR~104 by almost two magnitudes. This suggests substantial dust formation from SPIRITS~19q during its outburst.

Optical Keck/LRIS spectra of SPIRITS~19q were obtained within a year of the 2018 outburst peak. In addition to narrow emission lines likely associated with the nearby \Hii~region, the spectrum reveals clear detections of broad emission features from \Civ~and the \Ciii/\Heii~blend (Fig.~\ref{fig:LRIS}). These features are consistent with a WC star, and the absence of the lower excitation 5696-\AA\ C\,{\sc iii} line suggests an early WC sub-type \citep{Crowther1998}. SPIRITS~19q notably exhibits the same broad WC emission features as the WC4 star in N604-WRXc (Fig.~\ref{fig:HET}; \citealt{Garofali2019}). 
The observed continuum emission at $\lambda=5500$ \AA~in the LRIS spectrum of SPIRITS~19q corresponds to a $V$-band magnitude of $V\approx18.0$ and an absolute magnitude of $M_V\approx-9.5$. The absolute $V$-band magnitude exceeds that of the brightest supergiants in the LMC ($M_V\approx-8.5$; \citealt{Bonanos2009}), which indicates that SPIRITS~19q is most likely associated with a luminous stellar cluster. 

In Fig.~\ref{fig:LRISMod}, the LRIS spectrum of SPIRITS~19q is overlaid with the spectrum of the dust-free LMC WC4 star BAT99-11 (= Brey 10, HD 32402; \citealt{Torres1987, Neugent2018}) that has been scaled to the distance of NGC~2403\footnote{Assuming an LMC distance of 50 kpc} and combined with a power-law fit of the continuum emission from the 19q spectrum. The amplitude and profile of the WC features in the distance-scaled and continuum offset spectrum of BAT99-11 show a close agreement with the features from SPIRITS~19q. 
We therefore claim that SPIRITS~19q hosts a WC4 star.

The near-IR spectrum of SPIRITS~19q in Fig.~\ref{fig:NIRES} shows a narrow \Hei~feature combined with a broad component. This broad component appears consistent with the profile of the \Hei~feature exhibited by Galactic WR stars \citep{Howarth1992, Groh2007}. The detection of CO absorption bands as well as a possible CN absorption feature at around 1.1 $\mu$m suggests that cooler supergiants \citep{Rayner2009} are also present in the VS~51 \Hii~region.

Since SPIRITS~19q is likely associated with a star cluster in the VS~51 \Hii~region we must consider other possible origins of the 2018 outburst beyond episodic dust-formation from a WC4 binary system. We can rule out several possibilities based on its optical and near-IR spectra. Given the mid-IR magnitude and color of SPIRITS~19q, one of the most plausible origins of the outburst would be a highly obscured core-collapse supernova \citep{Jencson2019}. The optical and near-IR spectra, however, show no clear evidence of broadened ($\gtrsim2000$ km s$^{-1}$) hydrogen emission lines that are characteristic of core-collapse supernovae. SPIRITS~19q is therefore unlikely associated with an obscured supernova. The lack of broad hydrogen emission features also suggests SPIRITS~19q is not associated with a giant eruption from a luminous blue variable (LBV; e.g.~\citealt{Smith2014}). Other possible origins include a massive stellar merger like NGC 4490 OT2011 \citep{Smith2016} or an intermediate-luminosity red transient (ILRT) like NGC~300 OT2008 \citep{Bond2009}, but these events should be much rarer than a dust formation episode in a WC binary. We therefore consider SPIRITS~19q as a candidate episodic dust-forming WC system that exhibits efficient dust-formation.

\subsubsection{SPIRITS~16df}

SPIRITS~16df exhibits irregular and low-amplitude mid-IR variability in the \spitzer~observations that span 2004 Oct 7 - 2019 Aug 2 (MJD 53285 - 58697; Fig.~\ref{fig:IRLC}). The median mid-IR emission from SPIRITS~16df was F$_{3.6}=0.479\pm0.017$ mJy and F$_{4.5}=0.536\pm0.015$ mJy (M$_{[3.6]}=-13.09\pm0.04$ and M$_{[4.5]}=-13.70\pm0.03$), and the maximum amplitude of the observed variability from the median mid-IR emission was $\sim10$\% ($\sim0.1$ mag). 
The amplitude and timing of the variability is consistent at both 3.6 and 4.5 $\mu$m.
Extended emission surrounding SPIRITS~16df in the non-subtracted \spitzer~image (Fig.~\ref{fig:WCGal}) suggests it is associated with an \Hii~region (i.e.~VS~41).

Optical spectroscopy of SPIRITS~16df shown in Fig.~\ref{fig:LRIS} was obtained on 2020 Mar 23 (MJD 58931) with Keck/LRIS and shows the \Civ~and the \Ciii/\Heii~emission features. This LRIS spectrum of SPIRITS~16df is consistent with the optical spectrum of VS~41 presented by \citet{Drissen1999}.
The observed continuum emission from the LRIS spectrum of SPIRITS~16df at $\lambda=5500$ \AA~corresponds to a $V$-band magnitude of $V\approx18.2$ and an absolute magnitude of $M_V\approx-9.3$. Similar to the LRIS spectrum of SPIRITS~19q, the bright absolute $V$-band magnitude from the SPIRITS~16df spectrum indicates it is associated with a luminous stellar cluster.
HST/WFPC2 imaging of SPIRITS~16df's host \Hii~region by \citet{Drissen1999} indeed identify a compact cluster adjacent to the WR stars in VS~41.

The spectrum of SPIRITS~16df shows a strong resemblance to the WC features from SPIRITS~19q and SPIRITS~16ln/N604-WRXc \citep{Garofali2019}. In Fig.~\ref{fig:LRISMod}, the LRIS spectrum of SPIRITS~16df is overlaid with the distance-scaled spectrum of the LMC WC4 star BAT99-11 combined with a power-law fit of the continuum emission from 16df. The close agreement with the WC4 features from BAT99-11 indicates that SPIRITS~16df hosts a WC4 star. 

Since the mid-IR light curve of SPIRITS~16df does not exhibit obvious quiescent and outburst states like SPIRITS~19q or 16ln, the color magnitude diagram in Fig.~\ref{fig:CMD} only shows its median absolute [3.6] magnitude and median \spitzer~mid-IR color ($\scol=0.61\pm0.05$). This mid-IR color is consistent with emission from hot dust and is also similar to that of SPIRITS~19q, 16ln, and HD~36402 (Fig.~\ref{fig:CMD}). 
However, it is difficult to determine what fraction of the emission might originate from the WC star compared to the \Hii~region or other nearby stars.  
We consider SPIRITS~16df as a possible dust-forming WC binary, but further multi-wavelength monitoring observations will be crucial to confirm this hypothesis.

\begin{deluxetable*}{llllllll}
\tablecaption{Dust-Forming WC Candidate Properties}
\tablewidth{0.98\linewidth}
\tablehead{ID & Nearby \Hii/SF region & Ref. & Z/Z$_\odot$ & Ref. & WR Class & Var.~Type & Period $P$ (yr)}
\startdata
16ln & NGC 604 & Ga19 & 0.65 & Vi88, Ga19 & WC4 & Episodic &  $8.8>P>2.2$\\ 
19q  & NGC 2403 VS 51 & VS65 & 0.6 & Mo10 & WC4 & Episodic & $P\gtrsim12$ \\ 
16df  & NGC 2403  VS 41 & VS65 & 0.5 & Mo10 & WC4 & Irregular & ? \\ 
14apu & NGC5457+166.4+86.3 & Cr16 & 0.5 & Cr16 & WCE?  & Irregular & ? \\ 
18hb & NGC 6946-26 & Kh14 & 0.5 & Kh14 & WCE  & Episodic? & ? \\ 
14bqe & IC 1613 Reg 155 & Me09 & 0.16 & Br07 & ?  & Episodic &  $P\sim8.5$\\ 
\enddata
\tablecomments{Summary of dust-forming WC candidate properties including nearby \Hii/star-forming region in their host galaxy, environment metallicity, WR classification, mid-IR variability type (e.g.~\citealt{Williams2019}), and period estimate or constraints based on the \spitzer/IRAC light curve. Abbreviations correspond to the following references: Ga19 - \citet{Garofali2019}, Vi88 - \citet{Vilchez1988}, VS65 - \citet{Veron1965}, Mo10 - \citet{Moustakas2010}, Cr16 - \citet{Croxall2016}, Kh14 - \citet{Khramtsova2014}, Me09 - \citet{Melena2009}, Br07 - \citet{Bresolin2007}.}
\label{tab:WC}
\end{deluxetable*}

\subsection{Candidate Dust-Forming WC Stars in Messier~101, NGC~6946, and IC~1613: SPIRITS~14apu, SPIRITS~18hb, and SPIRITS~14bqe}

\subsubsection{SPIRITS~14apu}
SPIRITS~14apu was identified in the nearby spiral galaxy M101 (Fig.~\ref{fig:WCGal}) at $d=6.4$ Mpc ($\mu=29.04$; \citealt{Shappee2011}) and is located within $3\farcs7$ of the \Hii~region NGC5457+166.4+86.3 at a deprojected galactocentric radius of $\sim5.8$ kpc \citep{Croxall2016}. This \Hii~region exhibits an oxygen abundance of 12 + log(O/H) = $8.42\pm0.05$ \citep{Croxall2016}, which implies an LMC-like sub-solar metallicity environment with $\mathrm{Z}\sim0.5$ Z$_\odot$. \citet{Croxall2016} also indicate the presence of WR features in optical spectroscopy of this \Hii~region. Although SPIRITS~14apu may not be located within this \Hii~region given the $3\farcs7$ ($\sim120$ pc) offset, we assume that the surrounding environment of 14apu hosts a similar sub-solar metallicity.

\spitzer~observations of SPIRITS~14apu (Fig.~\ref{fig:IRLC}) revealed slow, irregular mid-IR variability with a prominent dip around late 2014/early 2015 (MJD $\sim57000$) followed by a gradual brightening that continued until the final observation taken on 2019 Oct 25 (MJD 58781).  At this final epoch, SPIRITS~14apu exhibited its maximum observed emission at 3.6 $\mu$m with F$_{3.6}=0.62\pm0.01$ mJy (M$_\mathrm{[3.6]}=-14.91\pm0.02$) and F$_{4.5}=0.71\pm0.01$ mJy (M$_\mathrm{[4.5]}=-15.53\pm0.01$). Given its bright absolute mid-IR magnitudes, SPIRITS~14apu is likely associated with a star cluster and/or dusty \Hii~region.

Two observations from 2004 Mar 8 (MJD 53072) captured SPIRITS~14apu at its minimum observed mid-IR brightness, where the average 3.6 and 4.5 $\mu$m emission was F$_{3.6}=0.40\pm0.01$ mJy (M$_\mathrm{[3.6]}=-14.43\pm0.03$) and F$_{4.5}=0.44\pm0.01$ mJy (M$_\mathrm{[4.5]}=-15.02\pm0.02$), respectively.
At this minimum brightness phase, SPIRITS~14apu exhibited a mid-IR color of $\scol=0.59\pm0.04$ (Fig.~\ref{fig:CMD}). It is assumed that the mid-IR emission at this observed minimum captures the non-variable emission from the nearby \Hii~region and/or cluster associated with SPIRITS~14apu. The cluster/\Hii~region-subtracted absolute [3.6] magnitude and mid-IR color of SPIRITS~14apu at its peak mid-IR brightness on 2019 Oct 25 is M$_\mathrm{[3.6]}=-13.58\pm0.08$ and $\scol=0.90\pm0.10$. Between the observed phases of minimum and maximum mid-IR emission, SPIRITS~14apu exhibited a red-ward shift in its mid-IR color consistent with the formation of hot dust. The absolute [3.6] magnitude of SPIRITS~14apu at its mid-IR peak was greater than WR~104 even after subtraction of the underlying cluster/\Hii~region emission, which suggests a high dust-formation rate relative to WR~104.

Spectroscopic observations of SPIRITS~14apu were obtained with Keck/LRIS on 2020 Mar 23 (MJD 58931) and show the \Ciii/\Heii~emission complex but present no significant detection of the \Civ~emission feature (Fig.~\ref{fig:LRIS}). 
The observed continuum emission at $\lambda=5500$ \AA~corresponds to a $V$-band magnitude of $V\approx19.1$ and an absolute magnitude of $M_V\approx-10.0$. This is consistent with the interpretation that SPIRITS~14apu is associated with a luminous stellar cluster.

Figure~\ref{fig:LRISMod} shows the LRIS spectrum of SPIRITS~14apu overlaid with spectrum of BAT99-11, which has been scaled to the distance of M101 and combined with a power-law fit of the continuum emission from 14apu. The width of the \Ciii/\Heii~emission complex in SPIRITS~14apu closely matches that of the distance-scaled and continuum-offset spectrum of BAT99-11. The non-detection of the \Civ~feature from 14apu is consistent with the low amplitude expected from this feature based on the spectrum of BAT99-11. 
We therefore consider SPIRITS~14apu to be a candidate dust-forming WC system. 
Although the \Ciii/\Heii~emission complex is indicative of a WC star, it is difficult to determine a specific WC classification for SPIRITS~14apu since the \Civ~feature is an important feature for distinguishing WC subtypes (e.g.~\citealt{Crowther1998}). Given the apparent link between early spectral sub-types and low metallicities (e.g.~\citealt{Crowther2007}), the sub-solar metallicity of its surrounding environment implies that SPIRITS~14apu likely hosts an early-type WC star.

\subsubsection{SPIRITS~18hb}
SPIRITS~18hb is located in a spiral arm $\sim2.8'$ from the central core of the galaxy NGC~6946  at a distance $d=7.72$ Mpc ($\mu=29.44$; \citealt{Anand2018}). An \Hii~region complex, NGC~6946-26, is located $3\farcs9$ from SPIRITS~18hb and exhibits an oxygen abundance of 12 + log(O/H) = $8.37\pm0.01$ \citep{Khramtsova2014}. This implies SPIRITS~18hb is in an LMC-like, sub-solar metallicity environment with $\mathrm{Z}\sim0.5$ Z$_\odot$.

\spitzer/IRAC observations of SPIRITS~18hb span from 2004 Jun 10 to 2019 Nov 8 (MJD 53167 - 58795) and show relatively steady mid-IR emission from 18hb until 2017 Nov 23 (MJD 58080; Fig.~\ref{fig:IRLC}). On 2017 Nov 23, the emission increased rapidly for $\sim1$ yr and had faded at both 3.6 and 4.5 $\mu$m in the last observation taken in 2019 Nov 8. This spike in mid-IR emission appears more consistent with episodic variability observed from WC candidates such as SPIRITS~16ln and 19q as opposed to the irregular variability observed from SPIRITS~16df and 14apu (Fig.~\ref{fig:IRLC}, Tab.~\ref{tab:WC})

The peak 3.6 $\mu$m emission from SPIRITS~18hb was observed on 2019 Feb 15 (MJD 58529) where F$_{3.6}=0.54\pm0.03$ mJy (M$_\mathrm{[3.6]}=-15.15\pm0.05$) and F$_{4.5}=0.50\pm0.02$ mJy (M$_\mathrm{[4.5]}=-15.56\pm0.05$). The median mid-IR emission from SPIRITS~18hb between 2004 Jun - 2017 Nov was F$_{3.6}=0.46\pm0.02$ mJy (M$_\mathrm{[3.6]}=-14.98\pm0.04$) and F$_{4.5}=0.38\pm0.02$ mJy (M$_\mathrm{[4.5]}=-15.24\pm0.06$). The bright mid-IR emission during this low-variability, quiescent phase and the extended emission around SPIRITS~18hb (Fig.~\ref{fig:WCGal}) indicates that 18hb is likely associated with a stellar cluster and/or dusty \Hii~region.

The mid-IR CMD in Fig.~\ref{fig:CMD} shows the red-ward shift of SPIRITS~18hb between its steady quiescent phase, where $\scol=0.27\pm0.05$, and its IR peak. 
The absolute [3.6] magnitude and mid-IR color at its IR peak subtracted by the quiescent phase emission is M$_\mathrm{[3.6]}=-13.14\pm0.42$ and $\scol=0.93\pm0.48$, respectively.
The shift in the mid-IR color from SPIRITS~18hb appears consistent with dust formation, but the uncertainties are large due to nearby extended emission and the crowded surrounding environment (Fig.~\ref{fig:WCGal}). Similar to SPIRITS~19q, 16df, and 14apu, the absolute [3.6] magnitude from SPIRITS~18hb during its IR peak exceeds that of the heavy dust-maker WR~104, which suggests a high dust-formation rate.

Optical spectra of SPIRITS~18hb were taken on 2019 Aug 20 (MJD 58715) with HET/LRS2 (Fig.~\ref{fig:HET}) and reveal detections of the \Civ~feature and \Ciii/\Heii~emission complex. Unfortunately, the \Ciii/\Heii~blend falls in the transition region between the UV and Orange channels of the LRS2 spectrograph which affects its emission profile. It is also difficult to compare the LRS2 spectra of SPIRITS~18hb to the WC4 star BAT99-11 (e.g.~Fig.~\ref{fig:LRISMod}) due to the non-photometric conditions of LRS2 spectra observations. However, the presence of these WC features and the absence of the lower excitation 5696-\AA\ C\,{\sc iii} emission line implies that SPIRITS~18hb hosts an early-type WC star. We therefore consider SPIRITS~18hb as a candidate dust-forming early-type WC system that exhibits episodic dust formation.

\subsubsection{SPIRITS~14bqe}
SPIRITS~14bqe is located in the nearby dwarf galaxy IC~1613 at a distance of 724 kpc ($\mu=24.30$; \citealt{Hatt2017}). The position of SPIRITS~14bqe is consistent with the star-forming region IC~1613 Region 155 identified by \citet{Melena2009}. IC~1613 hosts a low-metallicity environment where the mean oxygen abundance of massive stars and \Hii~regions is 12 + log(O/H) = $7.90\pm0.08$ \citep{Bresolin2007}. This implies SPIRITS~14bqe is in a low-metallicity environment with $\mathrm{Z}\sim0.16$ Z$_\odot$, which is the lowest metallicity of all 6 dust-forming WC candidates (Tab.~\ref{tab:WC}).

\spitzer/IRAC observations of SPIRITS~14bqe between 2010 Jan 26 and 2019 Nov 8 (MJD 55222 - 58795) show two mid-IR outburst peaks separated by $\sim8.5$ yr. The measured mid-IR emission was consistent at both mid-IR peaks, where F$_{3.6}=0.107\pm0.001$ mJy (M$_\mathrm{[3.6]}=-8.25\pm0.01$) and F$_{4.5}=0.189\pm0.001$ mJy (M$_\mathrm{[4.5]}=-9.35\pm0.01$) at the 2010 Feb 14 (MJD 55241) peak. The 2010 Feb mid-IR peak faded on a timescale of 2 years back to quiescence, where the median emission was F$_{3.6}=0.063\pm0.001$ mJy (M$_\mathrm{[3.6]}=-7.68\pm0.01$) and F$_{4.5}=0.082\pm0.001$ mJy (M$_\mathrm{[4.5]}=-8.45\pm0.01$). The absolute mid-IR magnitude exhibited by SPIRITS~14bqe during quiescence is consistent with emission from an individual stellar source as opposed to a stellar cluster or dusty \Hii~region.

Near-IR imaging of SPIRITS~14bqe taken on 2016 Oct 11 (MJD 57672) during mid-IR quiescence (Fig.~\ref{fig:IRLC}) with WIRC show a point-like IR counterpart at the \spitzer-derived coordinates of 14bqe (Fig.~\ref{fig:14bqe}). WIRC imaging also shows a nearby near-IR point source $1\farcs8$ south-east of the SPIRITS~14bqe that falls within the $2\farcs4$ aperture used for measuring \spitzer/IRAC photometry. However, \spitzer~imaging shows there is no significant mid-IR emission associated with the south-east near-IR point source. The \spitzer~photometry of SPIRITS~14bqe is therefore dominated by the emission from the central near-IR point source associated with 14bqe. The WIRC photometry measured from this near-IR counterpart is $J=18.94\pm0.04$ and $K_s=17.90\pm0.06$, which indicates absolute $J$ and $K_s$ magnitudes of M$_J=-5.36$ and M$_{Ks}=-6.40$. The average absolute $K_s$ magnitudes of Galactic non-dusty WC stars range from M$_{Ks}=$ -4.3 to -5.3, but for Galactic dusty WC9 stars the average absolute $K_s$ magnitude is M$_{Ks}=-6.6\pm0.8$  \citep{Rate2020}. The near-IR brightness from SPIRITS~14bqe therefore exceeds that of non-dusty WC stars but is consistent \rev{with the} brightness of dusty WC stars.

The mid-IR CMD in Fig.~\ref{fig:CMD} shows the red-ward shift from SPIRITS~14bqe between quiescence and its mid-IR peak, where $\scol = 0.77\pm0.02$ and $1.10\pm0.01$, respectively. Interestingly, the red mid-IR color from SPIRITS~14bqe during quiescence suggests it forms dust persistently and undergoes enhanced dust-formation episodes during the mid-IR outbursts. SPIRITS~14bqe is fainter than the peak emission from the persistent, variable dust-forming WC4 system HD~36402 but is brighter than the peak emission from the episodic, dust-forming WC5+O9 binary system WR~19.

\begin{figure}[t!]
    \centerline{\includegraphics[width=0.98\linewidth]{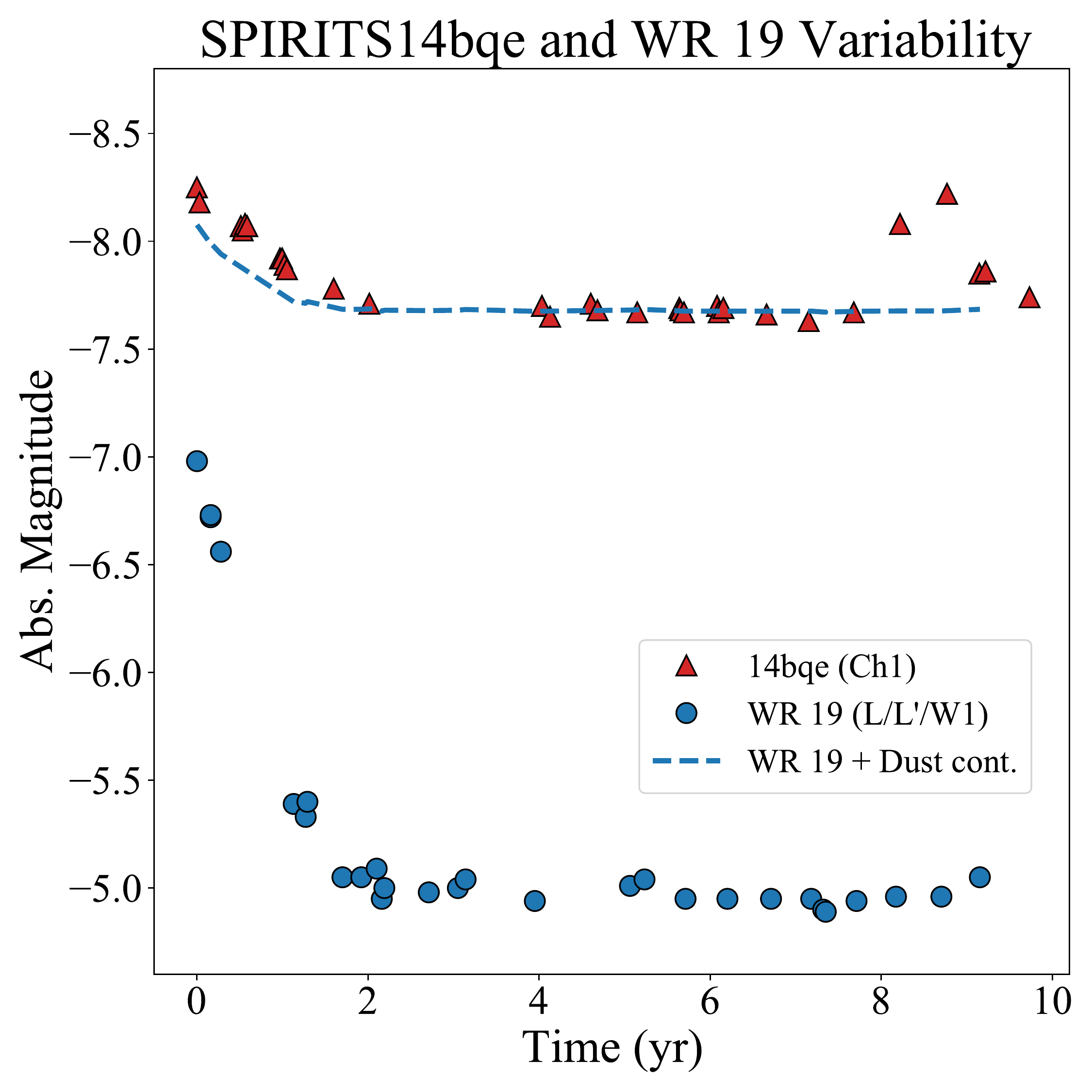}}
    \caption{Mid-IR light curve of SPIRITS~14bqe between 2010 Jan and 2019 Nov (MJD 55222 - 58796) plotted with the phase-folded mid-IR light curve of WR~19 over its 10.1 yr orbital period \citep{Williams1990b,Veen1998,Williams2019}. The origin $t = 0$ yr corresponds to the date of the maximum observed mid-IR emission from SPIRITS~14bqe and WR~19. The dashed line shows the WR~19 light curve with a constant flux offset added to match the quiescent mid-IR emission from SPIRITS~14bqe.}
    \label{fig:WR19}
\end{figure}

Figure~\ref{fig:WR19} shows a comparison of the absolute mid-IR magnitude light curves from SPIRITS~14bqe and WR~19 over the 10.1-yr orbital period of WR~19   \citep{Williams1990b,Veen1998,Williams2019}. 
Both systems show a similar decrease in mid-IR emission that returns to quiescence over a timescale of $\lesssim2$ yr following the outburst peak. 
Unlike SPIRITS~14bqe, the mid-IR emission from WR~19 is dominated by dust only during the periodic dust-formation events that correspond to periastron passage. During quiescence, mid-IR emission from WR~19 is largely due to the free-free emission from the stellar wind of the WC star \citep{Williams2019}. The absence of persistent dust formation in WR~19 is apparent in the discrepancy between the absolute mid-IR magnitudes of WR~19 and SPIRITS~14bqe in their quiescent phases. In quiescence, WR~19 notably exhibits a similar mid-IR color and absolute magnitude as the dust-free WC4 stars in the LMC (Fig.~\ref{fig:CMD}).

Figure~\ref{fig:WR19} also shows the mid-IR light curve of WR~19 with a constant flux offset added to match the quiescent mid-IR emission from SPIRITS~14bqe. This approximates the emission from persistent dust formation in addition to the observed periodic dust formation in WR~19. The dust-formation peak in this modified light curve of WR~19 closely resembles the observed light curve of SPIRITS~14bqe. It is therefore plausible that SPIRITS~14bqe is a WC system exhibiting persistent and episodic dust formation. However, the lack of optical spectroscopy precludes a firm classification of SPIRITS~14bqe.

\subsection{The Apparent Red WC4 Outlier in the LMC: HD~32125}
\label{Sec:Outlier}
In the mid-IR CMD (Fig.~\ref{fig:CMD}), there are two LMC WC4 sources that exhibit red $\scol$ colors consistent with dust emission: HD~36402 (= BAT99-38) and HD~32125 (= BAT99-9). HD~36402 is a known dusty WC system \citep{Williams2013}, but no evidence of circumstellar dust or mid-IR variability has been reported for HD~32125 \citep{Williams2019}. Upon closer inspection of the SAGE IRAC Epoch 1 and Epoch 2 catalog, the apparent [3.6] magnitudes of HD~32125 present discrepant values of $\mathrm{[3.6]}=14.05\pm0.09$ from Epoch 1 (2005 Jul) and $\mathrm{[3.6]}=13.55\pm0.05$ from Epoch 2 (2005 Oct/Nov); however, the apparent [4.5], [5.8], and [8.0] magnitudes are consistent between both Epochs. The average W1 ($\lambda_c=3.35$ $\mu$m) magnitude of HD~32125 reported by \citet{Williams2019}, who found no significant mid-IR variability from HD~32125, was $W1 = 13.52\pm0.03$, which is consistent with the [3.6] magnitude from SAGE Epoch 2. 

The [3.6] magnitude of HD~32125 in Fig.~\ref{fig:CMD} and from \citet{Bonanos2009} is identical to the discrepant Epoch 1 [3.6] magnitude reported in the SAGE IRAC Epoch 1 and Epoch 2 Catalog. The fainter [3.6] magnitude would account for its apparent red mid-IR color of $\scol=0.94$ in the CMD. If we instead adopt the SAGE Epoch 2 apparent [3.6] magnitude, the absolute [3.6] magnitude and mid-IR color of HD~32125 is M$_\mathrm{[3.6]}=-4.97$ and $\scol=0.45$. The SAGE Epoch 2 photometry of HD~32125 is therefore consistent with the dust-free WC4 stars in the LMC (Fig.~\ref{fig:CMD}). Given the observational evidence, we conclude that HD~32125 unlikely exhibited circumstellar dust emission and suggest that the fainter [3.6] measurement in SAGE Epoch 1 did not arise from any intrinsic astrophysical variability associated with HD~32125.

\begin{deluxetable*}{llllllll}
\tablecaption{\rev{SPIRITS~19q Dust Properties with Different Grain Assumptions}}
\tablewidth{0.98\linewidth}
\tablehead{\rule{0pt}{3ex} Grain Composition & $a$ ($\mu$m) & $T_\mathrm{d}$ (K) & $M_\mathrm{d}$ (M$_\odot$) & $\dot{M}_\mathrm{d}$ (M$_\odot$ yr$^{-1}$) & $\dot{M}_\mathrm{d}$/$\dot{M}_\mathrm{WR}$$^\mathrm{a}$ & $\langle \dot{M_d}\rangle^\mathrm{b}$ (M$_\odot$ yr$^{-1}$)}
\startdata
 Amorphous Carbon & 0.01 & $710^{+10}_{-10}$ & $6.3^{+0.4}_{-0.4} \times 10^{-6}$ & $6.4^{+0.4}_{-0.4} \times 10^{-6} $ & $30\%$ & $ \lesssim 5.2 \times 10^{-7} $\\
($\rho_b= 1.8$ g cm$^{-3}$) & 0.1 & $690^{+10}_{-10}$ & $6.6^{+0.4}_{-0.4} \times 10^{-6}$ & $6.7^{+0.4}_{-0.4} \times 10^{-6} $ & $30\%$ & $ \lesssim 5.5 \times 10^{-7} $\\
 & 0.5 & $680^{+10}_{-10}$ & $2.0^{+0.1}_{-0.1} \times 10^{-6}$ & $2.0^{+0.1}_{-0.1} \times 10^{-6} $ & $10\%$ & $ \lesssim 1.7 \times 10^{-7} $\\ \hline
Graphite & 0.01 & $680^{+10}_{-10}$ & $21.6^{+1.3}_{-1.4} \times 10^{-6}$ & $22.0^{+1.4}_{-1.5} \times 10^{-6} $ & $100\%$ & $ \lesssim 18.0 \times 10^{-7} $\\
($\rho_b= 2.2$ g cm$^{-3}$) & 0.1 & $650^{+10}_{-10}$ & $19.9^{+1.2}_{-1.3} \times 10^{-6}$ & $20.2^{+1.2}_{-1.3} \times 10^{-6} $ & $90\%$ & $ \lesssim 16.6 \times 10^{-7} $\\
 & 0.5 & $760^{+10}_{-10}$ & $2.7^{+0.2}_{-0.2} \times 10^{-6}$ & $2.8^{+0.2}_{-0.2} \times 10^{-6} $ & $10\%$ & $ \lesssim 2.3 \times 10^{-7} $\\ \hline
\enddata
\tablecomments{\rev{Derived SPIRITS~19q dust temperature ($T_d$), dust mass ($M_d$), dust production rate ($\dot{M}_d$), dust-to-gas mass fraction ($\dot{M}_\mathrm{d}$/$\dot{M}_\mathrm{WR}$), and the limit on the period-averaged dust input rate ($\langle \dot{M_d}\rangle$) assuming amorphous carbon or graphite compositions with a grain radius $a=0.01$, 0.1, and 0.5 $\mu$m. The amorphous carbon grain emissivity model is adopted from \citet{Zubko1996} and the graphite grain emissivity model is adopted from \citet{Draine2001} and \citet{Li2001}. The dust properties are determined from the quiescence-subtracted \spitzer~flux on 2018 Jun 20 (MJD 58289) of F$_{3.6}=1.393\pm0.013$ mJy and F$_{4.5}=1.655\pm0.008$ mJy. The dust production rate $\dot{M}_d$ is derived assuming the dust mass $M_d$ is produced over a $\Delta t=0.98$ yr timescale.\\a. The mean WC4 mass loss rate of $2.14\times10^{-5}$ M$_\odot$ yr$^{-1}$ derived by \citet{Sander2019} is adopted for $\dot{M}_\mathrm{WR}$.\\b. The periodicity of the dust formation event in SPIRITS~19q is assumed to be $\gtrsim12$ yr.}}
\label{tab:SPIRITS19q}
\end{deluxetable*}

\section{Discussion}
\label{sec:4}
\subsection{Efficient Dust-Formation from SPIRITS~19q}
\label{sec:Dust}

Out of all the six source in our sample, SPIRITS~19q exhibited the most luminous mid-IR emission at its IR peak after subtraction of the quiescent emission (Fig.~\ref{fig:CMD}). This suggests that SPIRITS~19q exhibits the most dust production of the sources in our sample. 
With the 3.6 and 4.5 $\mu$m \spitzer~photometry of SPIRITS~19q, we can estimate the dust production rate during its IR outburst and compare it against other known dust-forming WC systems. Since the mid-IR dust emission from SPIRITS~19q's IR outburst is significantly brighter than its quiescent emission, we are able to obtain more robust estimates of the dust properties associated with the outburst.

The hot dust mass around SPIRITS~19q can be estimated from the 3.6 and 4.5 $\mu$m flux measured by \spitzer. Assuming the mid-IR emission is optically thin, the mass of the emitting dust with a grain radius $a$ and temperature $T_d$ can be expressed as

\begin{equation}
M_d=\frac{(4/3)\,a\,\rho_b\,F_\nu\,d^2}{Q_d(\nu,a )\,B_\nu(T_d)},
\label{eq:Mass}
\end{equation}

\noindent
where $\rho_b$ is the dust grain bulk density, $F_\nu$ is the observed flux, $d$ is the distance to SPIRITS~19q, $Q_d(\nu,a)$ is the dust emission efficiency, and $B_\nu$ is the Planck function \rev{(See \citealt{Williams1990a})}. 
\rev{The dust temperature T$_d$ can be derived from the ratio of the measured fluxes at 3.6 and 4.5 $\mu$m, where the profile of the IR emission is defined by $F_\nu\propto B_\nu(T_d) \, Q_d(\nu,a)$. Observations of dust-formation from Galactic WC stars suggest that the newly formed dust can have a range of grain sizes from 0.01 $\mu$m to $\gtrsim 0.5$ $\mu$m \citep{Williams2009WR140,Rajagopal2007,Lau2020}. The C-rich winds of WC stars and observed featureless mid-IR continuum emission imply that WC dust grains are composed of amorphous carbon and/or graphite \citep{Cherchneff2000,Zubko2004,Gupta2020}. Given the range of WC dust properties, we estimate the dust properties from SPIRITS~19q assuming grain emissivity models for amorphous carbon \citep{Zubko1996} and graphite \citep{Draine2001,Li2001} with a uniform grain size radius of 0.01, 0.1, and 0.5 $\mu$m. }

The quiescence-subtracted \spitzer~flux during SPIRITS~19q's peak IR emission on 2018 Jun 20 (MJD 58289) was and F$_{3.6}=1.393\pm0.013$ mJy and F$_{4.5}=1.655\pm0.008$ mJy. %
The observed onset of the rising mid-IR emission on 2017 Jun 26 (MJD 57930) provides a temporal baseline, $\Delta t$, to estimate the dust production rate from SPIRITS~19q during its IR outburst. The time between the onset of the IR outburst and the IR peak is $\Delta t=0.98$ yr.
\rev{We estimate the dust production rates from SPIRITS~19q with the different grain assumptions based on these flux measurements and summarize the results in Table~\ref{tab:SPIRITS19q}. The SPIRITS~19q dust production rates range from $\dot{M}_d\approx(2 - 7)\times10^{-6}$ M$_\odot$ yr$^{-1}$ for an amorphous carbon composition and $\dot{M}_d\approx(3 - 22)\times10^{-6}$ M$_\odot$ yr$^{-1}$ for a graphite composition.} SPIRITS~19q's dust production rate during outburst \rev{may therefore exceed}
the most efficient Galactic WC dust-makers, WR~112 and WR~104, which exhibit persistent dust production rates of $3-4\times10^{-6}$ M$_\odot$ yr$^{-1}$ \citep{Lau2020,Lau2020b}. 

It is important to address that SPIRITS~19q exhibits episodic, as opposed to persistent, dust formation. 
\rev{Our observational constraints on the periodicity of dust-formation episodes from SPIRITS~19q provide a lower limit of $\gtrsim12$ yr (Tab.~\ref{tab:WC}). The total, period-averaged dust output rate, which is relevant to its contribution to the local dust budget, is therefore a factor of $\gtrsim12$ less than the dust production rate measured during the outburst (Tab.~\ref{tab:SPIRITS19q}). Although the episodic nature of dust-formation in SPIRITS~19q reduces the total dust input rate, this does not preclude our observations of enhanced, instantaneous dust-formation relative to known Galactic WC dust-makers.}

\rev{The outflow dust-to-gas mass fraction can be estimated for SPIRITS~19q} by adopting a WC wind mass-loss rate of $\dot{M}_\mathrm{\rev{WR}}=2.14\times10^{-5}$ M$_\odot$ yr$^{-1}$, which is consistent with the mean mass-loss rate derived from Galactic WC4 stars by \citet{Sander2019}. \rev{The derived dust production rates imply dust-to-gas mass fractions of $\sim10-30\%$ for amorphous carbon and $\sim10-100\%$ for graphite (Tab.~\ref{tab:SPIRITS19q}). Since it is highly unlikely that $90-100\%$ of the WC wind is forming dust, either the $a = 0.1$ and $0.01$ $\mu$m graphite dust models are not valid or the WC star in SPIRITS~19q exhibits a higher mass-loss rate than the mean WC4 mass-loss rate. However, even a $\sim10\%$ dust-to-gas mass fraction implies a high dust condensation efficiency given that LMC WC4 stars exhibit carbon surface abundances of $20-45\%$ \citep{Crowther2002}. } Interestingly, this challenges the observational evidence that suggests late-type WC stars (i.e.~WC8 and WC9) are the most efficient and heavy dust-forming WC systems \citep{Lau2020}, which may have major implications on the impact of the dust input from early-type WC stars.

\subsection{Colliding-wind Dust-formation in Early-type WC Binaries}

Dust-formation in early-type WC stars, which exhibit hotter temperatures and higher wind velocities than their late-type counterparts, is currently poorly understood given the dearth of known dust-forming WCE systems. Given that these early-type WC stars are likely the dominant population of WC stars in low-metallicity environments, it is important to understand their dust formation and mass-loss properties in order to characterize their impact on the dust abundance and chemistry of their host galaxy's ISM, especially in the early Universe. 
Based on the results of our dust-forming early-type WC candidates and our in-depth analysis on SPIRITS~19q, it is clear that the hot photospheres and high wind velocities from WC4 stars do not completely impede the formation of dust.

Due to higher wind velocities, early-type WC stars should exhibit reduced dust-formation efficiency relative to late-type WC stars based on the theoretical model of colliding-wind dust-formation in WR+OB binaries by \citet{Usov1991}. \citet{Usov1991} determine that $\alpha$, the fraction of the WR wind that is compressed and rapidly cooled in the wind-collision shock to form dust, can be characterized as

\begin{equation}
\alpha\propto \eta^2(1 + \eta^{1/2})^2 \dot{M}_\mathrm{WR}^2\,v_\mathrm{WR}^{-6}\,D^{-2},
\label{eq:Usov}
\end{equation}

\noindent
where $\eta$ is the momentum ratio between the OB and WR star winds $\left(\eta= \frac{\dot{M}_\mathrm{OB}v_\mathrm{OB}}{\dot{M}_\mathrm{WR}v_\mathrm{WR}}\right)$, $\dot{M}_\mathrm{WR/OB}$ is the mass-loss rate of the WR/OB star, $v_\mathrm{WR/OB}$ is the terminal velocity of the WR/OB star wind, and $D$ is the separation distance between the WR and OB stars. In this colliding-wind dust-formation framework, we will demonstrate that it is feasible for an early-type WC binary to achieve the same dust-formation efficiency as a late-type WC binary with an identical orbital separation $D$ if the OB companion in the early-type WC binary exhibits a higher mass-loss rate.
\rev{We note that clumping in the WC and OB-companion winds (e.g.~\citealt{Smith2014}) may also influence the colliding-wind dust-production efficiency; however, characterizing the effects of clumping is beyond the scope of this work.}

Although the average $\dot{M}_\mathrm{WR}$ derived for Galactic WC4 stars are consistent with that of Galactic WC9 stars ($2.2\times10^{-5}$ M$_\odot$ yr$^{-1}$; \citealt{Sander2019}), the terminal velocity of a WC4 wind ($v_\infty=3310$ km s$^{-1}$) exceeds that of a WC9 wind ($v_\infty=1390$ km s$^{-1}$) by more than factor of 2 \citep{Sander2019}.
For dust-forming WC4+OB and WC9+OB binaries with identical orbital separations, the wind momentum ratio $\eta$ must be enhanced in the WC4+OB system relative to the WC9+OB system by a factor of $\sim11$ in order to exhibit an equivalent dust formation efficiency (Eq.~\ref{eq:Usov}).

For example, $\eta\sim0.01$ in the dust-forming WC9+OB binary WR~104 \citep{Soulain2018}, where the OB-companion exhibits a mass-loss rate of $6\times10^{-8}$ M$_\odot$ yr$^{-1}$ \citep{Harries2004}. Therefore, a WC4+OB binary with the same $D$, $v_\mathrm{OB}$, and $\dot{M}_\mathrm{WR}$ as WR~104 would match its condensation efficiency if the OB-companion possessed a mass-loss rate of $\dot{M}_\mathrm{OB}\sim1.6\times10^{-6}$ M$_\odot$ yr$^{-1}$. This mass-loss rate is consistent with that of an O supergiant or an early O giant \citep{Muijres2012}.
\rev{Even at $1/3$ Z$_\odot$ metallicity, luminous O stars are predicted to exhibit mass-loss rates greater than $10^{-6}$ M$_\odot$ yr$^{-1}$ \citep{Vink2001}.} 
\rev{The dust-forming WC system }HD~36402 in the LMC notably hosts an O8 supergiant companion \citep{Moffat1990}. 
Not only is it plausible for a WC4 binary to host such a companion, but it is possible for \rev{an O supergiant companion} to possess \rev{a mass loss rate greater than $\sim1.6\times10^{-6}$ M$_\odot$ yr$^{-1}$} such that the dust condensation efficiency exceeds that of WR~104. We suggest that this is the reason for the efficient dust production rate measured from SPIRITS~19q (Sec.~\ref{sec:Dust}). A high mass-loss rate companion may also explain \rev{the} efficient dust formation from SPIRITS~16df, 14apu, and 18hb inferred from their high absolute [3.6] magnitudes (Fig.~\ref{fig:CMD}).

It is instructive to address the implications of the limit on the orbital period of SPIRITS~19q ($P\gtrsim12$ yr) on its dust-formation efficiency. This long orbital period implies the semi-major axis of its orbit should be larger than that of WR~104, which exhibits a near-circular orbit with a period of $P\approx0.66$ yr \citep{Tuthill2008}. If SPIRITS~19q had a circular orbit, this orbital period constraint should lead to a substantially lower dust-formation efficiency relative to WR~104 given that $\alpha\propto D^{-2}$ (Eq.~\ref{eq:Usov}). However, SPIRITS~19q's sharp IR outburst (Fig.~\ref{fig:IRLC}) indicates a highly eccentric orbit and thus a time-varying dust-formation efficiency. Dust-formation from SPIRITS~19q should therefore peak near periastron passage when the WC star is closest to its companion. This is the mechanism that we claim is responsible for its IR outburst. Orbital eccentricity is therefore an important parameter that regulates dust formation in long-period WC systems (e.g.~\citealt{Williams2009WR140}).

Individual WC systems that exhibit high dust formation rates could have a significant impact on the dust input rate in the ISM of galaxies, including those with low-metallicities. For example, the dust production rate \revv{estimates} of SPIRITS~19q during its outburst \revv{(Tab.~\ref{tab:SPIRITS19q}) range between the total dust production rates from the population of RSGs in the LMC ($\dot{M}_{d,RSG}=1.4\times10^{-6}$ M$_\odot$ yr$^{-1}$) to the population of AGB stars in the LMC ($\dot{M}_{d,AGB}=1.4\times10^{-5}$ M$_\odot$ yr$^{-1}$; \citealt{Riebel2012,Srinivasan2016}).}
Efficient dust production from WC binaries in LMC-like metallicities supports the results from the dust input and binary population synthesis models from \citet{Lau2020}. At LMC-like \rev{metallicities} with a constant star formation history, \citet{Lau2020} predicted that WC binaries input dust at a comparable rate to AGB stars within factors of a few. 

\section{Conclusions}
\label{sec:5}
We have presented \spitzer~mid-IR light curves of six extragalactic dust-forming WC candidates in low-metallicity environments from the SPIRITS survey (Fig.~\ref{fig:IRLC}). 
Over the $10+$ year temporal baselines of \spitzer/IRAC observations at 3.6 and 4.5 $\mu$m, these sources exhibited irregular or episodic mid-IR variability similar to that of known dust-forming WC systems (e.g.~\citealt{Williams2019}).
Follow-up optical spectroscopy with Keck/LRIS and HET/LRS2 of SPIRITS~16ln, 19q, 16df, 14apu, and 18hb confirmed emission from the \Civ~and/or the \Ciii/\Heii~blend consistent with early-type WC stars.
SPIRITS~14bqe, the only candidate without spectroscopic follow-up, exhibited periodic variability with repeating mid-IR emission peaks on $\sim8.5$ yr timescales. SPIRITS~14bqe's periodic mid-IR variability resembles that of the highly eccentric WC5+O9 binary system WR~19 (Fig.~\ref{fig:WR19}), which shows repeating dust formation and IR brightening episodes during periastron passage in its 10.1-yr orbit \citep{Williams2009WR19}.

We identified dust formation from the WC candidates based on their \spitzer~$\scol$ color. 
\spitzer/IRAC photometry of all candidates except for SPIRITS~14bqe, which is the most nearby candidate, appeared to suffer from confusion with nearby dusty \Hii~regions or stellar clusters. After subtracting the quiescent, non-variable emission, the mid-IR colors from SPIRITS~16ln, 19q, 14apu, and 18hb at their mid-IR peak were consistent with dust formation (Fig.~\ref{fig:CMD}). During quiescence and outburst, the mid-IR colors of SPIRITS~14bqe were consistent with the presence and formation of dust. Although the irregular mid-IR variability from SPIRITS~16df made it difficult to define and subtract the quiescent mid-IR emission component, its median mid-IR color was consistent with the presence of hot dust.

We presented SPIRITS~16ln as the variable mid-IR counterpart of the WC4+O system N604-WRXc, which also exhibits X-ray variability consistent with a colliding-wind binary \citep{Garofali2019}. This colliding-wind mechanism is associated with dust-formation in WC+OB binaries, which supports our interpretation of SPIRITS~16ln's mid-IR outburst as a dust-formation episode. Based on the spectroscopic classification, X-ray variability, and mid-IR variability, we claimed that SPIRITS~16ln/N604-WRXc is a dust-forming colliding-wind WC binary. SPIRITS~16ln/N604-WRXc is the therefore first dust-forming colliding-wind WC binary to be identified beyond the LMC.

The absolute [3.6] magnitudes of SPIRITS~19q, 16df, 14apu, and 18hb exceeded those of one of the brightest known dusty WC systems, WR~104 (Fig.~\ref{fig:CMD}), which suggests these candidates may be exhibiting highly efficient dust formation for a WC binary. SPIRITS~19q exhibited a luminous mid-IR outburst that was significantly brighter than its mid-IR emission in quiescence. We investigated the dust properties from SPIRITS~19q during outburst using the 3.6 and 4.5 $\mu$m emission measured from \spitzer. 
\rev{We derived dust production rates for SPIRITS~19q ranging from $\dot{M}_d\approx(2 - 7)\times10^{-6}$ M$_\odot$ yr$^{-1}$ for an amorphous carbon composition and $\dot{M}_d\approx(3 - 22)\times10^{-6}$ M$_\odot$ yr$^{-1}$ for a graphite composition (Tab.~\ref{tab:SPIRITS19q}). The SPIRITS~19q dust production rate during its outburst may therefore exceed that of} the most efficient known dust-forming WC systems ($\dot{M}_d\approx3-4\times10^{-6}$ M$_\odot$ yr$^{-1}$; \citealt{Lau2020,Lau2020b}).

Based on the theoretical framework of dust production in colliding-wind WR-OB binaries by \citet{Usov1991}, we show that efficient dust-formation from early-type WC binaries is plausible with a high mass-loss rate companion such as an O supergiant or early O giant star. We therefore suggest that early-type WC systems, which are more common than late-type WC systems in low-metallicity environments, may indeed have a significant impact on the dust input in the ISM of low-metallicity galaxies. This supports the model results from \citet{Lau2020} that show the dust production rate the population of WC binaries is comparable to that of the AGB star population at LMC-like metallicities with constant star formation histories. 

Our results underscore the value of investigating dust formation from early-type WC systems in low-metallicity environments. 
We also emphasize the importance of the continued exploration of dust production from WC systems within our Galaxy, where spatially resolved observations allow us to study the interplay between dust formation and the orbital and mass-loss properties of the central WC binary (e.g.~\citealt{Callingham2019,Han2020,Lau2020b}). Spatially resolved observations of dust formed by WC binaries are also crucial for probing how the newly-formed dust evolves as it expands and integrates into the surrounding ISM (e.g.~\citealt{Marchenko2002,Williams2009WR140}).
Lastly, our results demonstrate the necessity of on-going long-term IR photometric monitoring programs like the recently-commissioned Palomar Gattini-IR \citep{De2020} for detecting episodic variations that would otherwise be missed.

\acknowledgments
\textit{Acknowledgements.}
We acknowledge S.~Anand and C.~Fremling for their support on the observations obtained with Keck I/LRIS.
We thank N.~Morrell, P.~Mudumba, and R.~Gehrz for their valuable feedback on our study. 
\rev{We also thank the anonymous referee for their insightful review and comments that have improved the quality of our work.}
RML acknowledges the Japan Aerospace Exploration Agency's International Top Young Fellowship (ITYF).
AFJM is grateful for financial assistance from NSERC (Canada).
This work is based on observations and archival data obtained with the Spitzer Space Telescope, which was operated by the Jet Propulsion Laboratory, California Institute of Technology under a contract with NASA. Support for this work was provided by NASA through an award issued by JPL/Caltech.
Some of the data presented herein were obtained at the W. M. Keck Observatory, which is operated as a scientific partnership among the California Institute of Technology, the University of California and the National Aeronautics and Space Administration. The Observatory was made possible by the generous financial support of the W. M. Keck Foundation.
The authors wish to recognize and acknowledge the very significant cultural role and reverence that the summit of Maunakea has always had within the indigenous Hawaiian community.  We are most fortunate to have the opportunity to conduct observations from this mountain.
The Hobby-Eberly Telescope (HET) is a joint project of the University of Texas at Austin, the Pennsylvania State University, Ludwig-Maximilians-Universit\rev{\"a}t M\rev{\"u}nchen, and Georg-August-\rev{Universit\"at} \rev{G\"ottingen}. The HET is named in honor of its principal benefactors, William P. Hobby and Robert E. Eberly.
The Low Resolution Spectrograph 2 (LRS2) was developed and funded by the University of Texas at Austin McDonald Observatory and Department of Astronomy and by Pennsylvania State University. We thank the Leibniz-Institut \rev{f\"ur} Astrophysik Potsdam (AIP) and the Institut \rev{f\"ur} Astrophysik \rev{G\"ottingen} (IAG) for their contributions to the construction of the integral field units.
Based partially on observations made with the NASA/ESA Hubble Space Telescope, obtained from the data archive at the Space Telescope Science Institute. STScI is operated by the Association of Universities for Research in Astronomy, Inc. under NASA contract NAS 5-26555.


%

\vspace{5mm}
\facilities{Spitzer(IRAC), Keck:I (LRIS), Keck:II (NIRES), HET(LRS2), P200(WIRC), HST(ACS/WFC)}


\software{ 
          \rev{ Astropy \citep{Astropy2013, Astropy2018},
          LPipe \citep{Perley2019},
          Spex-tool \citep{Cushing2004},
          Panacea (\url{https://github.com/grzeimann/Panacea})
          }
          }

\newpage

\refstepcounter{table}
\setcounter{table}{4}
\startlongtable
\begin{deluxetable*}{lllccllll}
\tablecaption{\textit{Spitzer}/IRAC Photometry of SPIRITS Dust-Forming WC Candidates}
\tablewidth{0pt}
\tablehead{MJD & F$_\mathrm{3.6}$ (mJy) & F$_\mathrm{4.5}$ (mJy) & [3.6] & [4.5] & [3.6] - [4.5] & M$_\mathrm{[3.6]}$ & M$_\mathrm{[4.5]}$ & PI (Prog ID)}
\startdata
\textbf{SPIRITS16ln}\\ 
53001.39 & 0.632 $\pm$ 0.030 & 0.600 $\pm$ 0.026 & 14.12 $\pm$ 0.05 & 13.69 $\pm$ 0.05 & 0.43 $\pm$ 0.07 & -10.50 & -10.93 & Houck (63)\\
53013.72 & 0.615 $\pm$ 0.029 & 0.593 $\pm$ 0.026 & 14.15 $\pm$ 0.05 & 13.70 $\pm$ 0.05 & 0.45 $\pm$ 0.07 & -10.47 & -10.92 & Gehrz (5)\\
53208.96 & 0.646 $\pm$ 0.030 & 0.624 $\pm$ 0.026 & 14.10 $\pm$ 0.05 & 13.65 $\pm$ 0.05 & 0.45 $\pm$ 0.07 & -10.52 & -10.97 & Gehrz (5)\\
53233.18 & 0.646 $\pm$ 0.030 & 0.603 $\pm$ 0.026 & 14.10 $\pm$ 0.05 & 13.69 $\pm$ 0.05 & 0.41 $\pm$ 0.07 & -10.52 & -10.93 & Gehrz (5)\\
53391.68 & 0.618 $\pm$ 0.029 & 0.597 $\pm$ 0.026 & 14.14 $\pm$ 0.05 & 13.70 $\pm$ 0.05 & 0.45 $\pm$ 0.07 & -10.48 & -10.92 & Gehrz (5)\\
53607.18 & 0.645 $\pm$ 0.030 & 0.595 $\pm$ 0.026 & 14.10 $\pm$ 0.05 & 13.70 $\pm$ 0.05 & 0.40 $\pm$ 0.07 & -10.52 & -10.92 & Gehrz (5)\\
53770.66 & 0.621 $\pm$ 0.030 & 0.573 $\pm$ 0.026 & 14.14 $\pm$ 0.05 & 13.74 $\pm$ 0.05 & 0.40 $\pm$ 0.07 & -10.48 & -10.88 & Gehrz (5)\\
57111.70 & 0.973 $\pm$ 0.029 & 0.979 $\pm$ 0.026 & 13.65 $\pm$ 0.03 & 13.16 $\pm$ 0.03 & 0.49 $\pm$ 0.04 & -10.97 & -11.46 & Kasliwal (11063)\\
57118.84 & 0.969 $\pm$ 0.029 & 0.971 $\pm$ 0.025 & 13.66 $\pm$ 0.03 & 13.17 $\pm$ 0.03 & 0.49 $\pm$ 0.04 & -10.96 & -11.45 & Kasliwal (11063)\\
57139.32 & 0.952 $\pm$ 0.028 & 0.942 $\pm$ 0.025 & 13.68 $\pm$ 0.03 & 13.20 $\pm$ 0.03 & 0.47 $\pm$ 0.04 & -10.94 & -11.42 & Kasliwal (11063)\\
57314.62 & 0.798 $\pm$ 0.029 & 0.777 $\pm$ 0.025 & 13.87 $\pm$ 0.04 & 13.41 $\pm$ 0.04 & 0.46 $\pm$ 0.05 & -10.75 & -11.21 & Kasliwal (11063)\\
57322.09 & 0.809 $\pm$ 0.029 & 0.780 $\pm$ 0.025 & 13.85 $\pm$ 0.04 & 13.41 $\pm$ 0.04 & 0.45 $\pm$ 0.05 & -10.77 & -11.21 & Kasliwal (11063)\\
57336.48 & 0.780 $\pm$ 0.029 & 0.770 $\pm$ 0.025 & 13.89 $\pm$ 0.04 & 13.42 $\pm$ 0.04 & 0.47 $\pm$ 0.05 & -10.73 & -11.20 & Kasliwal (11063)\\
57491.30 & 0.711 $\pm$ 0.029 & 0.682 $\pm$ 0.026 & 13.99 $\pm$ 0.04 & 13.55 $\pm$ 0.04 & 0.44 $\pm$ 0.06 & -10.63 & -11.07 & Kasliwal (11063)\\\hline
\textbf{SPIRITS19q}\\
53285.07 & 0.353 $\pm$ 0.009 & 0.324 $\pm$ 0.008 & 14.75 $\pm$ 0.03 & 14.36 $\pm$ 0.03 & 0.39 $\pm$ 0.04 & -12.76 & -13.15 & Van Dyk (226)\\
53286.42 & 0.346 $\pm$ 0.009 & 0.328 $\pm$ 0.008 & 14.77 $\pm$ 0.03 & 14.35 $\pm$ 0.03 & 0.42 $\pm$ 0.04 & -12.74 & -13.16 & Kennicutt (159)\\
53290.08 & 0.346 $\pm$ 0.009 & 0.330 $\pm$ 0.009 & 14.77 $\pm$ 0.03 & 14.34 $\pm$ 0.03 & 0.43 $\pm$ 0.04 & -12.74 & -13.17 & Kennicutt (159)\\
53310.11 & 0.364 $\pm$ 0.008 & 0.334 $\pm$ 0.008 & 14.72 $\pm$ 0.03 & 14.33 $\pm$ 0.03 & 0.39 $\pm$ 0.04 & -12.79 & -13.18 & Van Dyk (226)\\
53453.94 & 0.340 $\pm$ 0.009 & 0.311 $\pm$ 0.008 & 14.79 $\pm$ 0.03 & 14.40 $\pm$ 0.03 & 0.39 $\pm$ 0.04 & -12.72 & -13.11 & Van Dyk (226)\\
53663.71 & 0.347 $\pm$ 0.008 & 0.323 $\pm$ 0.008 & 14.77 $\pm$ 0.03 & 14.36 $\pm$ 0.03 & 0.41 $\pm$ 0.04 & -12.74 & -13.15 & Meikle (20256)\\
53817.89 & 0.343 $\pm$ 0.008 & 0.318 $\pm$ 0.008 & 14.78 $\pm$ 0.03 & 14.38 $\pm$ 0.03 & 0.40 $\pm$ 0.04 & -12.73 & -13.13 & Meikle (20256)\\
54036.16 & 0.337 $\pm$ 0.009 & 0.319 $\pm$ 0.008 & 14.80 $\pm$ 0.03 & 14.38 $\pm$ 0.03 & 0.42 $\pm$ 0.04 & -12.71 & -13.13 & Meikle (30292)\\
54039.12 & 0.306 $\pm$ 0.012 & 0.303 $\pm$ 0.014 & 14.91 $\pm$ 0.04 & 14.43 $\pm$ 0.05 & 0.48 $\pm$ 0.07 & -12.60 & -13.08 & Sugerman (30494)\\
54192.83 & 0.330 $\pm$ 0.008 & 0.307 $\pm$ 0.008 & 14.82 $\pm$ 0.03 & 14.42 $\pm$ 0.03 & 0.40 $\pm$ 0.04 & -12.69 & -13.09 & Sugerman (30494)\\
54192.84 & 0.328 $\pm$ 0.008 & 0.310 $\pm$ 0.008 & 14.83 $\pm$ 0.03 & 14.41 $\pm$ 0.03 & 0.42 $\pm$ 0.04 & -12.68 & -13.10 & Meikle (30292)\\
54423.33 & 0.336 $\pm$ 0.008 & 0.325 $\pm$ 0.008 & 14.80 $\pm$ 0.03 & 14.36 $\pm$ 0.03 & 0.44 $\pm$ 0.04 & -12.71 & -13.15 & Sugerman (30494)\\
54427.14 & 0.342 $\pm$ 0.008 & 0.326 $\pm$ 0.008 & 14.79 $\pm$ 0.03 & 14.35 $\pm$ 0.03 & 0.44 $\pm$ 0.04 & -12.72 & -13.16 & Meixner (40010)\\
54428.01 & 0.337 $\pm$ 0.009 & 0.326 $\pm$ 0.008 & 14.80 $\pm$ 0.03 & 14.35 $\pm$ 0.03 & 0.45 $\pm$ 0.04 & -12.71 & -13.16 & Kotak (40619)\\
54563.79 & 0.341 $\pm$ 0.008 & 0.329 $\pm$ 0.008 & 14.79 $\pm$ 0.03 & 14.34 $\pm$ 0.03 & 0.45 $\pm$ 0.04 & -12.72 & -13.17 & Kotak (40619)\\
54568.21 & 0.342 $\pm$ 0.008 & 0.321 $\pm$ 0.008 & 14.79 $\pm$ 0.03 & 14.37 $\pm$ 0.03 & 0.42 $\pm$ 0.04 & -12.72 & -13.14 & Meixner (40010)\\
55149.91 & 0.336 $\pm$ 0.009 & -- $\pm$ -- & 14.81 $\pm$ 0.03 & -- $\pm$ -- & -- $\pm$ -- & -12.70 & -- & Freedman (61002)\\
55168.68 & 0.345 $\pm$ 0.008 & -- $\pm$ -- & 14.78 $\pm$ 0.03 & -- $\pm$ -- & -- $\pm$ -- & -12.73 & -- & Freedman (61002)\\
55185.89 & 0.351 $\pm$ 0.008 & -- $\pm$ -- & 14.76 $\pm$ 0.03 & -- $\pm$ -- & -- $\pm$ -- & -12.75 & -- & Freedman (61002)\\
55302.67 & 0.344 $\pm$ 0.008 & -- $\pm$ -- & 14.78 $\pm$ 0.03 & -- $\pm$ -- & -- $\pm$ -- & -12.73 & -- & Freedman (61002)\\
55339.50 & 0.350 $\pm$ 0.008 & -- $\pm$ -- & 14.76 $\pm$ 0.03 & -- $\pm$ -- & -- $\pm$ -- & -12.75 & -- & Freedman (61002)\\
55890.71 & 0.283 $\pm$ 0.013 & 0.318 $\pm$ 0.008 & 14.99 $\pm$ 0.05 & 14.38 $\pm$ 0.03 & 0.61 $\pm$ 0.06 & -12.52 & -13.13 & Kochanek (80015)\\
56092.54 & 0.356 $\pm$ 0.009 & 0.342 $\pm$ 0.007 & 14.74 $\pm$ 0.03 & 14.30 $\pm$ 0.02 & 0.44 $\pm$ 0.04 & -12.77 & -13.21 & Kochanek (80015)\\
56671.43 & 0.358 $\pm$ 0.009 & 0.320 $\pm$ 0.008 & 14.74 $\pm$ 0.03 & 14.37 $\pm$ 0.03 & 0.37 $\pm$ 0.04 & -12.77 & -13.14 & Kasliwal (10136)\\
56795.32 & 0.341 $\pm$ 0.008 & 0.319 $\pm$ 0.008 & 14.79 $\pm$ 0.03 & 14.38 $\pm$ 0.03 & 0.41 $\pm$ 0.04 & -12.72 & -13.13 & Kasliwal (10136)\\
56827.40 & 0.356 $\pm$ 0.008 & 0.331 $\pm$ 0.008 & 14.74 $\pm$ 0.03 & 14.34 $\pm$ 0.03 & 0.40 $\pm$ 0.04 & -12.77 & -13.17 & Kasliwal (10136)\\
57061.94 & 0.346 $\pm$ 0.008 & 0.325 $\pm$ 0.008 & 14.77 $\pm$ 0.03 & 14.36 $\pm$ 0.03 & 0.41 $\pm$ 0.04 & -12.74 & -13.15 & Kasliwal (11063)\\
57174.60 & 0.353 $\pm$ 0.008 & 0.332 $\pm$ 0.008 & 14.75 $\pm$ 0.03 & 14.33 $\pm$ 0.03 & 0.42 $\pm$ 0.04 & -12.76 & -13.18 & Kasliwal (11063)\\
57180.80 & 0.336 $\pm$ 0.009 & 0.309 $\pm$ 0.009 & 14.80 $\pm$ 0.03 & 14.41 $\pm$ 0.03 & 0.39 $\pm$ 0.04 & -12.71 & -13.10 & Kasliwal (11063)\\
57201.79 & 0.338 $\pm$ 0.008 & 0.351 $\pm$ 0.008 & 14.80 $\pm$ 0.03 & 14.27 $\pm$ 0.03 & 0.53 $\pm$ 0.04 & -12.71 & -13.24 & Kasliwal (11063)\\
57388.85 & 0.370 $\pm$ 0.008 & 0.334 $\pm$ 0.008 & 14.70 $\pm$ 0.02 & 14.33 $\pm$ 0.03 & 0.37 $\pm$ 0.04 & -12.81 & -13.18 & Kasliwal (11063)\\
57395.55 & 0.363 $\pm$ 0.009 & 0.326 $\pm$ 0.008 & 14.72 $\pm$ 0.03 & 14.35 $\pm$ 0.03 & 0.37 $\pm$ 0.04 & -12.79 & -13.16 & Kasliwal (11063)\\
57409.70 & 0.361 $\pm$ 0.008 & 0.329 $\pm$ 0.008 & 14.73 $\pm$ 0.03 & 14.34 $\pm$ 0.03 & 0.39 $\pm$ 0.04 & -12.78 & -13.17 & Kasliwal (11063)\\
57763.83 & 0.364 $\pm$ 0.009 & 0.336 $\pm$ 0.008 & 14.72 $\pm$ 0.03 & 14.32 $\pm$ 0.03 & 0.40 $\pm$ 0.04 & -12.79 & -13.19 & Kasliwal (13053)\\
57929.60 & 0.428 $\pm$ 0.009 & 0.408 $\pm$ 0.009 & 14.54 $\pm$ 0.02 & 14.11 $\pm$ 0.02 & 0.43 $\pm$ 0.03 & -12.97 & -13.40 & Kasliwal (13053)\\
58129.33 & 1.443 $\pm$ 0.010 & 1.516 $\pm$ 0.009 & 13.22 $\pm$ 0.01 & 12.68 $\pm$ 0.01 & 0.54 $\pm$ 0.01 & -14.29 & -14.83 & Kasliwal (13053)\\
58288.90 & 1.738 $\pm$ 0.010 & 1.980 $\pm$ 0.009 & 13.02 $\pm$ 0.01 & 12.39 $\pm$ 0.01 & 0.63 $\pm$ 0.01 & -14.49 & -15.12 & Kasliwal (13053)\\
58507.74 & 1.527 $\pm$ 0.010 & 1.791 $\pm$ 0.009 & 13.16 $\pm$ 0.01 & 12.50 $\pm$ 0.01 & 0.66 $\pm$ 0.01 & -14.35 & -15.01 & Kasliwal (14089)\\
58549.28 & 1.523 $\pm$ 0.010 & 1.693 $\pm$ 0.008 & 13.16 $\pm$ 0.01 & 12.56 $\pm$ 0.01 & 0.60 $\pm$ 0.01 & -14.35 & -14.95 & Kasliwal (14089)\\
58655.40 & 1.277 $\pm$ 0.010 & 1.528 $\pm$ 0.010 & 13.36 $\pm$ 0.01 & 12.68 $\pm$ 0.01 & 0.68 $\pm$ 0.01 & -14.15 & -14.83 & Kasliwal (14089)\\
58697.23 & 1.180 $\pm$ 0.010 & 1.449 $\pm$ 0.009 & 13.44 $\pm$ 0.01 & 12.73 $\pm$ 0.01 & 0.71 $\pm$ 0.01 & -14.07 & -14.78 & Kasliwal (14089)\\ \hline
\textbf{SPIRITS16df}\\
53285.07 & 0.498 $\pm$ 0.018 & 0.543 $\pm$ 0.014 & 14.38 $\pm$ 0.04 & 13.80 $\pm$ 0.03 & 0.58 $\pm$ 0.05 & -13.13 & -13.71 & Van Dyk (226)\\
53286.42 & 0.471 $\pm$ 0.017 & 0.515 $\pm$ 0.015 & 14.44 $\pm$ 0.04 & 13.86 $\pm$ 0.03 & 0.58 $\pm$ 0.05 & -13.07 & -13.65 & Kennicutt (159)\\
53290.08 & 0.484 $\pm$ 0.017 & 0.529 $\pm$ 0.015 & 14.41 $\pm$ 0.04 & 13.83 $\pm$ 0.03 & 0.58 $\pm$ 0.05 & -13.10 & -13.68 & Kennicutt (159)\\
53310.11 & 0.476 $\pm$ 0.018 & 0.538 $\pm$ 0.014 & 14.43 $\pm$ 0.04 & 13.81 $\pm$ 0.03 & 0.62 $\pm$ 0.05 & -13.08 & -13.70 & Van Dyk (226)\\
53453.94 & 0.461 $\pm$ 0.017 & 0.502 $\pm$ 0.014 & 14.46 $\pm$ 0.04 & 13.89 $\pm$ 0.03 & 0.57 $\pm$ 0.05 & -13.05 & -13.62 & Van Dyk (226)\\
53663.71 & 0.475 $\pm$ 0.017 & 0.556 $\pm$ 0.014 & 14.43 $\pm$ 0.04 & 13.77 $\pm$ 0.03 & 0.66 $\pm$ 0.05 & -13.08 & -13.74 & Meikle (20256)\\
53817.89 & 0.466 $\pm$ 0.017 & 0.531 $\pm$ 0.014 & 14.45 $\pm$ 0.04 & 13.82 $\pm$ 0.03 & 0.63 $\pm$ 0.05 & -13.06 & -13.69 & Meikle (20256)\\
54036.16 & 0.465 $\pm$ 0.017 & 0.540 $\pm$ 0.015 & 14.45 $\pm$ 0.04 & 13.81 $\pm$ 0.03 & 0.64 $\pm$ 0.05 & -13.06 & -13.70 & Meikle (30292)\\
54039.12 & 0.429 $\pm$ 0.019 & 0.512 $\pm$ 0.020 & 14.54 $\pm$ 0.05 & 13.86 $\pm$ 0.04 & 0.68 $\pm$ 0.06 & -12.97 & -13.65 & Sugerman (30494)\\
54192.83 & 0.491 $\pm$ 0.018 & 0.536 $\pm$ 0.014 & 14.39 $\pm$ 0.04 & 13.81 $\pm$ 0.03 & 0.58 $\pm$ 0.05 & -13.12 & -13.70 & Sugerman (30494)\\
54192.84 & 0.484 $\pm$ 0.017 & 0.530 $\pm$ 0.014 & 14.41 $\pm$ 0.04 & 13.83 $\pm$ 0.03 & 0.58 $\pm$ 0.05 & -13.10 & -13.68 & Meikle (30292)\\
54423.33 & 0.469 $\pm$ 0.017 & 0.533 $\pm$ 0.014 & 14.44 $\pm$ 0.04 & 13.82 $\pm$ 0.03 & 0.62 $\pm$ 0.05 & -13.07 & -13.69 & Sugerman (30494)\\
54427.14 & 0.476 $\pm$ 0.017 & 0.538 $\pm$ 0.015 & 14.43 $\pm$ 0.04 & 13.81 $\pm$ 0.03 & 0.62 $\pm$ 0.05 & -13.08 & -13.70 & Meixner (40010)\\
54428.01 & 0.462 $\pm$ 0.017 & 0.525 $\pm$ 0.014 & 14.46 $\pm$ 0.04 & 13.83 $\pm$ 0.03 & 0.63 $\pm$ 0.05 & -13.05 & -13.68 & Kotak (40619)\\
54563.79 & 0.476 $\pm$ 0.018 & 0.511 $\pm$ 0.015 & 14.43 $\pm$ 0.04 & 13.87 $\pm$ 0.03 & 0.56 $\pm$ 0.05 & -13.08 & -13.64 & Kotak (40619)\\
54568.21 & 0.479 $\pm$ 0.017 & 0.545 $\pm$ 0.014 & 14.42 $\pm$ 0.04 & 13.79 $\pm$ 0.03 & 0.63 $\pm$ 0.05 & -13.09 & -13.72 & Meixner (40010)\\
55890.71 & 0.470 $\pm$ 0.018 & 0.533 $\pm$ 0.013 & 14.44 $\pm$ 0.04 & 13.82 $\pm$ 0.03 & 0.62 $\pm$ 0.05 & -13.07 & -13.69 & Kochanek (80015)\\
56092.54 & 0.538 $\pm$ 0.017 & 0.572 $\pm$ 0.014 & 14.29 $\pm$ 0.03 & 13.74 $\pm$ 0.03 & 0.55 $\pm$ 0.04 & -13.22 & -13.77 & Kochanek (80015)\\
56650.66 & 0.471 $\pm$ 0.018 & 0.524 $\pm$ 0.014 & 14.44 $\pm$ 0.04 & 13.84 $\pm$ 0.03 & 0.60 $\pm$ 0.05 & -13.07 & -13.67 & Kochanek (10001)\\
56671.43 & 0.492 $\pm$ 0.017 & 0.537 $\pm$ 0.015 & 14.39 $\pm$ 0.04 & 13.81 $\pm$ 0.03 & 0.58 $\pm$ 0.05 & -13.12 & -13.70 & Kasliwal (10136)\\
56795.32 & 0.466 $\pm$ 0.017 & 0.511 $\pm$ 0.014 & 14.45 $\pm$ 0.04 & 13.87 $\pm$ 0.03 & 0.58 $\pm$ 0.05 & -13.06 & -13.64 & Kasliwal (10136)\\
56827.40 & 0.515 $\pm$ 0.017 & 0.539 $\pm$ 0.014 & 14.34 $\pm$ 0.04 & 13.81 $\pm$ 0.03 & 0.53 $\pm$ 0.05 & -13.17 & -13.70 & Kasliwal (10136)\\
57061.94 & 0.531 $\pm$ 0.018 & 0.579 $\pm$ 0.014 & 14.31 $\pm$ 0.04 & 13.73 $\pm$ 0.03 & 0.58 $\pm$ 0.05 & -13.20 & -13.78 & Kasliwal (11063)\\
57174.60 & 0.490 $\pm$ 0.017 & 0.528 $\pm$ 0.015 & 14.40 $\pm$ 0.04 & 13.83 $\pm$ 0.03 & 0.57 $\pm$ 0.05 & -13.11 & -13.68 & Kasliwal (11063)\\
57180.80 & 0.519 $\pm$ 0.017 & 0.562 $\pm$ 0.014 & 14.33 $\pm$ 0.04 & 13.76 $\pm$ 0.03 & 0.57 $\pm$ 0.04 & -13.18 & -13.75 & Kasliwal (11063)\\
57201.79 & 0.525 $\pm$ 0.018 & 0.574 $\pm$ 0.015 & 14.32 $\pm$ 0.04 & 13.74 $\pm$ 0.03 & 0.58 $\pm$ 0.05 & -13.19 & -13.77 & Kasliwal (11063)\\
57388.85 & 0.512 $\pm$ 0.017 & 0.562 $\pm$ 0.014 & 14.35 $\pm$ 0.04 & 13.76 $\pm$ 0.03 & 0.59 $\pm$ 0.05 & -13.16 & -13.75 & Kasliwal (11063)\\
57395.55 & 0.494 $\pm$ 0.017 & 0.543 $\pm$ 0.014 & 14.39 $\pm$ 0.04 & 13.80 $\pm$ 0.03 & 0.59 $\pm$ 0.05 & -13.12 & -13.71 & Kasliwal (11063)\\
57409.70 & 0.499 $\pm$ 0.017 & 0.573 $\pm$ 0.015 & 14.38 $\pm$ 0.04 & 13.74 $\pm$ 0.03 & 0.64 $\pm$ 0.05 & -13.13 & -13.77 & Kasliwal (11063)\\
57763.83 & 0.446 $\pm$ 0.017 & 0.510 $\pm$ 0.015 & 14.50 $\pm$ 0.04 & 13.87 $\pm$ 0.03 & 0.63 $\pm$ 0.05 & -13.01 & -13.64 & Kasliwal (13053)\\
57929.60 & 0.482 $\pm$ 0.017 & 0.541 $\pm$ 0.015 & 14.41 $\pm$ 0.04 & 13.80 $\pm$ 0.03 & 0.61 $\pm$ 0.05 & -13.10 & -13.71 & Kasliwal (13053)\\
58129.33 & 0.480 $\pm$ 0.017 & 0.536 $\pm$ 0.015 & 14.42 $\pm$ 0.04 & 13.81 $\pm$ 0.03 & 0.61 $\pm$ 0.05 & -13.09 & -13.70 & Kasliwal (13053)\\
58288.90 & 0.446 $\pm$ 0.017 & 0.486 $\pm$ 0.015 & 14.50 $\pm$ 0.04 & 13.92 $\pm$ 0.03 & 0.58 $\pm$ 0.05 & -13.01 & -13.59 & Kasliwal (13053)\\
58507.74 & 0.494 $\pm$ 0.017 & 0.543 $\pm$ 0.014 & 14.39 $\pm$ 0.04 & 13.80 $\pm$ 0.03 & 0.59 $\pm$ 0.05 & -13.12 & -13.71 & Kasliwal (14089)\\
58549.28 & 0.494 $\pm$ 0.018 & 0.549 $\pm$ 0.014 & 14.39 $\pm$ 0.04 & 13.79 $\pm$ 0.03 & 0.60 $\pm$ 0.05 & -13.12 & -13.72 & Kasliwal (14089)\\
58655.40 & 0.466 $\pm$ 0.017 & 0.492 $\pm$ 0.015 & 14.45 $\pm$ 0.04 & 13.91 $\pm$ 0.03 & 0.54 $\pm$ 0.05 & -13.06 & -13.60 & Kasliwal (14089)\\
58697.23 & 0.455 $\pm$ 0.018 & 0.513 $\pm$ 0.015 & 14.48 $\pm$ 0.04 & 13.86 $\pm$ 0.03 & 0.62 $\pm$ 0.05 & -13.03 & -13.65 & Kasliwal (14089)\\ \hline
\textbf{SPIRITS14apu}\\
53072.09 & 0.395 $\pm$ 0.010 & 0.430 $\pm$ 0.009 & 14.63 $\pm$ 0.03 & 14.05 $\pm$ 0.02 & 0.58 $\pm$ 0.04 & -14.41 & -14.99 & Rieke (60)\\
53072.49 & 0.408 $\pm$ 0.009 & 0.458 $\pm$ 0.008 & 14.59 $\pm$ 0.02 & 13.98 $\pm$ 0.02 & 0.61 $\pm$ 0.03 & -14.45 & -15.06 & Rieke (60)\\
55960.72 & 0.549 $\pm$ 0.009 & -- $\pm$ -- & 14.27 $\pm$ 0.02 & -- $\pm$ -- & -- $\pm$ -- & -14.77 & -- & Kasliwal (80196)\\
55980.99 & 0.556 $\pm$ 0.009 & -- $\pm$ -- & 14.26 $\pm$ 0.02 & -- $\pm$ -- & -- $\pm$ -- & -14.78 & -- & Kasliwal (80196)\\
56165.01 & -- $\pm$ -- & 0.612 $\pm$ 0.008 & -- $\pm$ -- & 13.67 $\pm$ 0.01 & -- $\pm$ -- & -- & -15.37 & Garnavich (80126)\\
56337.07 & 0.563 $\pm$ 0.010 & -- $\pm$ -- & 14.25 $\pm$ 0.02 & -- $\pm$ -- & -- $\pm$ -- & -14.79 & -- & Kasliwal (90240)\\
56348.11 & 0.565 $\pm$ 0.010 & -- $\pm$ -- & 14.24 $\pm$ 0.02 & -- $\pm$ -- & -- $\pm$ -- & -14.80 & -- & Garnavich (80126)\\
56393.76 & 0.614 $\pm$ 0.010 & -- $\pm$ -- & 14.15 $\pm$ 0.02 & -- $\pm$ -- & -- $\pm$ -- & -14.89 & -- & Kasliwal (90240)\\
56516.35 & -- $\pm$ -- & 0.645 $\pm$ 0.008 & -- $\pm$ -- & 13.61 $\pm$ 0.01 & -- $\pm$ -- & -- & -15.43 & Kasliwal (90240)\\
56742.84 & 0.508 $\pm$ 0.009 & 0.617 $\pm$ 0.007 & 14.36 $\pm$ 0.02 & 13.66 $\pm$ 0.01 & 0.70 $\pm$ 0.02 & -14.68 & -15.38 & Kasliwal (10136)\\
56771.83 & 0.512 $\pm$ 0.009 & 0.622 $\pm$ 0.007 & 14.35 $\pm$ 0.02 & 13.65 $\pm$ 0.01 & 0.70 $\pm$ 0.02 & -14.69 & -15.39 & Kasliwal (10136)\\
56902.01 & 0.443 $\pm$ 0.010 & 0.572 $\pm$ 0.008 & 14.50 $\pm$ 0.02 & 13.74 $\pm$ 0.02 & 0.76 $\pm$ 0.03 & -14.54 & -15.30 & Kasliwal (10136)\\
57136.69 & 0.442 $\pm$ 0.009 & 0.546 $\pm$ 0.008 & 14.51 $\pm$ 0.02 & 13.79 $\pm$ 0.02 & 0.72 $\pm$ 0.03 & -14.53 & -15.25 & Kasliwal (11063)\\
57144.06 & 0.425 $\pm$ 0.009 & 0.522 $\pm$ 0.008 & 14.55 $\pm$ 0.02 & 13.84 $\pm$ 0.02 & 0.71 $\pm$ 0.03 & -14.49 & -15.20 & Kasliwal (11063)\\
57150.17 & 0.454 $\pm$ 0.010 & 0.549 $\pm$ 0.008 & 14.48 $\pm$ 0.02 & 13.79 $\pm$ 0.02 & 0.69 $\pm$ 0.03 & -14.56 & -15.25 & Kasliwal (11063)\\
57163.71 & 0.457 $\pm$ 0.009 & 0.558 $\pm$ 0.008 & 14.47 $\pm$ 0.02 & 13.77 $\pm$ 0.02 & 0.70 $\pm$ 0.03 & -14.57 & -15.27 & Kasliwal (11063)\\
57191.82 & 0.465 $\pm$ 0.009 & 0.574 $\pm$ 0.008 & 14.45 $\pm$ 0.02 & 13.74 $\pm$ 0.01 & 0.71 $\pm$ 0.03 & -14.59 & -15.30 & Kasliwal (11063)\\
57220.79 & 0.461 $\pm$ 0.009 & 0.585 $\pm$ 0.007 & 14.46 $\pm$ 0.02 & 13.72 $\pm$ 0.01 & 0.74 $\pm$ 0.02 & -14.58 & -15.32 & Kasliwal (11063)\\
57247.82 & 0.439 $\pm$ 0.009 & 0.561 $\pm$ 0.007 & 14.51 $\pm$ 0.02 & 13.76 $\pm$ 0.01 & 0.75 $\pm$ 0.03 & -14.53 & -15.28 & Kasliwal (11063)\\
57486.85 & 0.487 $\pm$ 0.009 & 0.632 $\pm$ 0.007 & 14.40 $\pm$ 0.02 & 13.63 $\pm$ 0.01 & 0.77 $\pm$ 0.02 & -14.64 & -15.41 & Kasliwal (11063)\\
57843.93 & 0.552 $\pm$ 0.010 & 0.648 $\pm$ 0.008 & 14.27 $\pm$ 0.02 & 13.61 $\pm$ 0.01 & 0.66 $\pm$ 0.02 & -14.77 & -15.43 & Kasliwal (13053)\\
57926.90 & 0.585 $\pm$ 0.010 & 0.713 $\pm$ 0.009 & 14.20 $\pm$ 0.02 & 13.50 $\pm$ 0.01 & 0.70 $\pm$ 0.02 & -14.84 & -15.54 & Kasliwal (13053)\\
58009.67 & 0.509 $\pm$ 0.009 & 0.609 $\pm$ 0.008 & 14.35 $\pm$ 0.02 & 13.68 $\pm$ 0.01 & 0.67 $\pm$ 0.02 & -14.69 & -15.36 & Kasliwal (13053)\\
58232.95 & 0.557 $\pm$ 0.009 & 0.657 $\pm$ 0.007 & 14.26 $\pm$ 0.02 & 13.59 $\pm$ 0.01 & 0.67 $\pm$ 0.02 & -14.78 & -15.45 & Kasliwal (13053)\\
58292.87 & 0.550 $\pm$ 0.010 & 0.666 $\pm$ 0.008 & 14.27 $\pm$ 0.02 & 13.58 $\pm$ 0.01 & 0.69 $\pm$ 0.02 & -14.77 & -15.46 & Kasliwal (13053)\\
58380.22 & 0.573 $\pm$ 0.009 & 0.685 $\pm$ 0.008 & 14.23 $\pm$ 0.02 & 13.55 $\pm$ 0.01 & 0.68 $\pm$ 0.02 & -14.81 & -15.49 & Kasliwal (13053)\\
58572.08 & 0.529 $\pm$ 0.009 & 0.644 $\pm$ 0.008 & 14.31 $\pm$ 0.02 & 13.61 $\pm$ 0.01 & 0.70 $\pm$ 0.02 & -14.73 & -15.43 & Kasliwal (14089)\\
58614.39 & 0.523 $\pm$ 0.009 & 0.633 $\pm$ 0.008 & 14.33 $\pm$ 0.02 & 13.63 $\pm$ 0.01 & 0.70 $\pm$ 0.02 & -14.71 & -15.41 & Kasliwal (14089)\\
58655.68 & 0.587 $\pm$ 0.009 & 0.682 $\pm$ 0.008 & 14.20 $\pm$ 0.02 & 13.55 $\pm$ 0.01 & 0.65 $\pm$ 0.02 & -14.84 & -15.49 & Kasliwal (14089)\\
58697.50 & 0.585 $\pm$ 0.010 & 0.701 $\pm$ 0.008 & 14.20 $\pm$ 0.02 & 13.52 $\pm$ 0.01 & 0.68 $\pm$ 0.02 & -14.84 & -15.52 & Kasliwal (14089)\\
58740.01 & 0.533 $\pm$ 0.009 & 0.648 $\pm$ 0.008 & 14.30 $\pm$ 0.02 & 13.61 $\pm$ 0.01 & 0.69 $\pm$ 0.02 & -14.74 & -15.43 & Kasliwal (14089)\\
58781.31 & 0.624 $\pm$ 0.010 & 0.711 $\pm$ 0.008 & 14.13 $\pm$ 0.02 & 13.51 $\pm$ 0.01 & 0.62 $\pm$ 0.02 & -14.91 & -15.53 & Kasliwal (14089)\\ \hline
\textbf{SPIRITS18hb}\\
53166.76 & 0.428 $\pm$ 0.027 & 0.355 $\pm$ 0.022 & 14.54 $\pm$ 0.07 & 14.26 $\pm$ 0.07 & 0.28 $\pm$ 0.10 & -14.90 & -15.18 & Kennicutt (159)\\
53260.29 & -- $\pm$ -- & 0.335 $\pm$ 0.022 & -- $\pm$ -- & 14.32 $\pm$ 0.07 & -- $\pm$ -- & -- & -15.12 & Meikle (3248)\\
53334.74 & 0.388 $\pm$ 0.026 & 0.328 $\pm$ 0.022 & 14.65 $\pm$ 0.07 & 14.35 $\pm$ 0.07 & 0.30 $\pm$ 0.10 & -14.79 & -15.09 & Kennicutt (159)\\
53571.24 & -- $\pm$ -- & 0.318 $\pm$ 0.021 & -- $\pm$ -- & 14.38 $\pm$ 0.07 & -- $\pm$ -- & -- & -15.06 & Meikle (20256)\\
53630.81 & -- $\pm$ -- & 0.337 $\pm$ 0.021 & -- $\pm$ -- & 14.32 $\pm$ 0.07 & -- $\pm$ -- & -- & -15.12 & Sugerman (20320)\\
53676.05 & -- $\pm$ -- & 0.364 $\pm$ 0.022 & -- $\pm$ -- & 14.23 $\pm$ 0.06 & -- $\pm$ -- & -- & -15.21 & Meikle (20256)\\
53734.89 & -- $\pm$ -- & 0.334 $\pm$ 0.025 & -- $\pm$ -- & 14.33 $\pm$ 0.08 & -- $\pm$ -- & -- & -15.11 & Sugerman (20320)\\
54065.89 & -- $\pm$ -- & 0.361 $\pm$ 0.022 & -- $\pm$ -- & 14.24 $\pm$ 0.07 & -- $\pm$ -- & -- & -15.20 & Meikle (30292)\\
54098.03 & -- $\pm$ -- & 0.367 $\pm$ 0.021 & -- $\pm$ -- & 14.23 $\pm$ 0.06 & -- $\pm$ -- & -- & -15.21 & Sugerman (30494)\\
54284.97 & 0.419 $\pm$ 0.026 & -- $\pm$ -- & 14.56 $\pm$ 0.07 & -- $\pm$ -- & -- $\pm$ -- & -14.88 & -- & Sugerman (30494)\\
54461.07 & -- $\pm$ -- & 0.400 $\pm$ 0.021 & -- $\pm$ -- & 14.13 $\pm$ 0.06 & -- $\pm$ -- & -- & -15.31 & Meixner (40010)\\
54492.85 & -- $\pm$ -- & 0.380 $\pm$ 0.022 & -- $\pm$ -- & 14.19 $\pm$ 0.06 & -- $\pm$ -- & -- & -15.25 & Kotak (40619)\\
54665.79 & 0.459 $\pm$ 0.025 & -- $\pm$ -- & 14.47 $\pm$ 0.06 & -- $\pm$ -- & -- $\pm$ -- & -14.97 & -- & Meixner (40010)\\
55049.01 & 0.400 $\pm$ 0.024 & -- $\pm$ -- & 14.62 $\pm$ 0.07 & -- $\pm$ -- & -- $\pm$ -- & -14.82 & -- & Andrews (60071)\\
55201.62 & -- $\pm$ -- & 0.373 $\pm$ 0.021 & -- $\pm$ -- & 14.21 $\pm$ 0.06 & -- $\pm$ -- & -- & -15.23 & Andrews (60071)\\
55421.58 & 0.438 $\pm$ 0.025 & -- $\pm$ -- & 14.52 $\pm$ 0.06 & -- $\pm$ -- & -- $\pm$ -- & -14.92 & -- & Andrews (70008)\\
55531.43 & -- $\pm$ -- & 0.374 $\pm$ 0.020 & -- $\pm$ -- & 14.21 $\pm$ 0.06 & -- $\pm$ -- & -- & -15.23 & Kochanek (70040)\\
55566.90 & -- $\pm$ -- & 0.376 $\pm$ 0.021 & -- $\pm$ -- & 14.20 $\pm$ 0.06 & -- $\pm$ -- & -- & -15.24 & Andrews (70008)\\
55769.87 & 0.455 $\pm$ 0.026 & -- $\pm$ -- & 14.48 $\pm$ 0.06 & -- $\pm$ -- & -- $\pm$ -- & -14.96 & -- & Kochanek (80015)\\
55774.83 & 0.461 $\pm$ 0.026 & -- $\pm$ -- & 14.46 $\pm$ 0.06 & -- $\pm$ -- & -- $\pm$ -- & -14.98 & -- & Andrews (80131)\\
55939.12 & -- $\pm$ -- & 0.379 $\pm$ 0.021 & -- $\pm$ -- & 14.19 $\pm$ 0.06 & -- $\pm$ -- & -- & -15.25 & Andrews (80131)\\
56521.15 & 0.459 $\pm$ 0.026 & -- $\pm$ -- & 14.47 $\pm$ 0.06 & -- $\pm$ -- & -- $\pm$ -- & -14.97 & -- & Andrews (90178)\\
56660.59 & -- $\pm$ -- & 0.392 $\pm$ 0.021 & -- $\pm$ -- & 14.15 $\pm$ 0.06 & -- $\pm$ -- & -- & -15.29 & Kochanek (10081)\\
56705.62 & -- $\pm$ -- & 0.383 $\pm$ 0.021 & -- $\pm$ -- & 14.18 $\pm$ 0.06 & -- $\pm$ -- & -- & -15.26 & Sugerman (10002)\\
56742.56 & 0.461 $\pm$ 0.026 & 0.403 $\pm$ 0.021 & 14.46 $\pm$ 0.06 & 14.12 $\pm$ 0.06 & 0.34 $\pm$ 0.08 & -14.98 & -15.32 & Kasliwal (10136)\\
56916.20 & 0.468 $\pm$ 0.026 & 0.394 $\pm$ 0.021 & 14.45 $\pm$ 0.06 & 14.15 $\pm$ 0.06 & 0.30 $\pm$ 0.08 & -14.99 & -15.29 & Kasliwal (10136)\\
56945.57 & 0.481 $\pm$ 0.026 & 0.396 $\pm$ 0.021 & 14.42 $\pm$ 0.06 & 14.14 $\pm$ 0.06 & 0.28 $\pm$ 0.08 & -15.02 & -15.30 & Kasliwal (10136)\\
57293.99 & 0.446 $\pm$ 0.023 & 0.376 $\pm$ 0.020 & 14.50 $\pm$ 0.06 & 14.20 $\pm$ 0.06 & 0.30 $\pm$ 0.08 & -14.94 & -15.24 & Kasliwal (11063)\\
57322.05 & 0.495 $\pm$ 0.024 & 0.435 $\pm$ 0.022 & 14.38 $\pm$ 0.05 & 14.04 $\pm$ 0.05 & 0.34 $\pm$ 0.08 & -15.06 & -15.40 & Kasliwal (11063)\\
57351.45 & 0.438 $\pm$ 0.024 & 0.447 $\pm$ 0.021 & 14.52 $\pm$ 0.06 & 14.01 $\pm$ 0.05 & 0.51 $\pm$ 0.08 & -14.92 & -15.43 & Kasliwal (11063)\\
57673.30 & 0.490 $\pm$ 0.027 & 0.432 $\pm$ 0.022 & 14.39 $\pm$ 0.06 & 14.05 $\pm$ 0.06 & 0.34 $\pm$ 0.08 & -15.05 & -15.39 & Kasliwal (13053)\\
58080.79 & 0.499 $\pm$ 0.026 & 0.467 $\pm$ 0.022 & 14.38 $\pm$ 0.06 & 13.96 $\pm$ 0.05 & 0.42 $\pm$ 0.08 & -15.06 & -15.48 & Krafton (13239)\\
58178.15 & 0.508 $\pm$ 0.026 & 0.470 $\pm$ 0.021 & 14.36 $\pm$ 0.06 & 13.96 $\pm$ 0.05 & 0.40 $\pm$ 0.07 & -15.08 & -15.48 & Krafton (13239)\\
58381.85 & 0.524 $\pm$ 0.026 & 0.493 $\pm$ 0.022 & 14.32 $\pm$ 0.05 & 13.90 $\pm$ 0.05 & 0.42 $\pm$ 0.07 & -15.12 & -15.54 & Krafton (13239)\\
58404.83 & 0.528 $\pm$ 0.024 & 0.471 $\pm$ 0.022 & 14.31 $\pm$ 0.05 & 13.95 $\pm$ 0.05 & 0.36 $\pm$ 0.07 & -15.13 & -15.49 & Fox (14098)\\
58449.30 & 0.507 $\pm$ 0.025 & 0.495 $\pm$ 0.022 & 14.36 $\pm$ 0.05 & 13.90 $\pm$ 0.05 & 0.46 $\pm$ 0.07 & -15.08 & -15.54 & Kasliwal (14089)\\
58515.57 & 0.533 $\pm$ 0.026 & 0.508 $\pm$ 0.023 & 14.31 $\pm$ 0.05 & 13.87 $\pm$ 0.05 & 0.44 $\pm$ 0.07 & -15.13 & -15.57 & Krafton (13239)\\
58529.84 & 0.542 $\pm$ 0.027 & 0.504 $\pm$ 0.022 & 14.29 $\pm$ 0.05 & 13.88 $\pm$ 0.05 & 0.41 $\pm$ 0.07 & -15.15 & -15.56 & Kasliwal (14089)\\
58607.95 & 0.537 $\pm$ 0.025 & 0.536 $\pm$ 0.021 & 14.30 $\pm$ 0.05 & 13.81 $\pm$ 0.04 & 0.49 $\pm$ 0.07 & -15.14 & -15.63 & Kasliwal (14089)\\
58795.40 & 0.463 $\pm$ 0.026 & 0.444 $\pm$ 0.022 & 14.46 $\pm$ 0.06 & 14.02 $\pm$ 0.05 & 0.44 $\pm$ 0.08 & -14.98 & -15.42 & Kasliwal (14089)\\ \hline
\textbf{SPIRITS14bqe}\\
52994.85 & -- $\pm$ -- & 0.115 $\pm$ 0.001 & -- $\pm$ -- & 15.48 $\pm$ 0.01 & -- $\pm$ -- & -- & -8.82 & Gehrz (128)\\
53727.66 & 0.070 $\pm$ 0.001 & -- $\pm$ -- & 16.51 $\pm$ 0.02 & -- $\pm$ -- & -- $\pm$ -- & -7.79 & -- & Gehrz (128)\\
55222.14 & 0.087 $\pm$ 0.001 & 0.149 $\pm$ 0.001 & 16.28 $\pm$ 0.01 & 15.21 $\pm$ 0.01 & 1.07 $\pm$ 0.01 & -8.02 & -9.09 & Freedman (61001)\\
55233.99 & 0.100 $\pm$ 0.001 & 0.182 $\pm$ 0.001 & 16.12 $\pm$ 0.01 & 14.98 $\pm$ 0.01 & 1.14 $\pm$ 0.01 & -8.18 & -9.32 & Freedman (61001)\\
55241.02 & 0.107 $\pm$ 0.001 & 0.189 $\pm$ 0.001 & 16.05 $\pm$ 0.01 & 14.95 $\pm$ 0.01 & 1.10 $\pm$ 0.01 & -8.25 & -9.35 & Freedman (61001)\\
55252.40 & 0.100 $\pm$ 0.001 & 0.172 $\pm$ 0.001 & 16.12 $\pm$ 0.01 & 15.05 $\pm$ 0.01 & 1.07 $\pm$ 0.01 & -8.18 & -9.25 & Freedman (61001)\\
55428.67 & 0.090 $\pm$ 0.001 & 0.147 $\pm$ 0.001 & 16.23 $\pm$ 0.01 & 15.22 $\pm$ 0.01 & 1.01 $\pm$ 0.01 & -8.07 & -9.08 & Freedman (61001)\\
55436.71 & 0.089 $\pm$ 0.001 & 0.151 $\pm$ 0.001 & 16.25 $\pm$ 0.02 & 15.19 $\pm$ 0.01 & 1.06 $\pm$ 0.01 & -8.05 & -9.11 & Freedman (61001)\\
55446.80 & 0.092 $\pm$ 0.001 & 0.143 $\pm$ 0.001 & 16.22 $\pm$ 0.02 & 15.25 $\pm$ 0.01 & 0.97 $\pm$ 0.01 & -8.08 & -9.05 & Freedman (61001)\\
55456.01 & 0.091 $\pm$ 0.001 & 0.146 $\pm$ 0.001 & 16.23 $\pm$ 0.02 & 15.22 $\pm$ 0.01 & 1.01 $\pm$ 0.01 & -8.07 & -9.08 & Freedman (61001)\\
55595.93 & 0.079 $\pm$ 0.001 & 0.120 $\pm$ 0.001 & 16.38 $\pm$ 0.02 & 15.44 $\pm$ 0.01 & 0.94 $\pm$ 0.02 & -7.92 & -8.86 & Freedman (61001)\\
55605.12 & 0.079 $\pm$ 0.001 & 0.113 $\pm$ 0.001 & 16.38 $\pm$ 0.02 & 15.50 $\pm$ 0.01 & 0.88 $\pm$ 0.02 & -7.92 & -8.80 & Freedman (61001)\\
55615.06 & 0.076 $\pm$ 0.001 & 0.117 $\pm$ 0.001 & 16.41 $\pm$ 0.02 & 15.46 $\pm$ 0.01 & 0.95 $\pm$ 0.02 & -7.89 & -8.84 & Freedman (61001)\\
55625.55 & 0.075 $\pm$ 0.001 & 0.115 $\pm$ 0.001 & 16.43 $\pm$ 0.02 & 15.48 $\pm$ 0.01 & 0.95 $\pm$ 0.02 & -7.87 & -8.82 & Freedman (61001)\\
55825.25 & 0.070 $\pm$ 0.001 & 0.104 $\pm$ 0.001 & 16.52 $\pm$ 0.02 & 15.60 $\pm$ 0.01 & 0.92 $\pm$ 0.02 & -7.78 & -8.70 & Boyer (80063)\\
55977.41 & 0.065 $\pm$ 0.001 & 0.095 $\pm$ 0.001 & 16.59 $\pm$ 0.02 & 15.69 $\pm$ 0.01 & 0.90 $\pm$ 0.02 & -7.71 & -8.61 & Boyer (80063)\\
56714.37 & 0.064 $\pm$ 0.001 & -- $\pm$ -- & 16.60 $\pm$ 0.02 & -- $\pm$ -- & -- $\pm$ -- & -7.70 & -- & Colbert (90230)\\
56748.93 & 0.062 $\pm$ 0.001 & 0.089 $\pm$ 0.001 & 16.65 $\pm$ 0.02 & 15.77 $\pm$ 0.01 & 0.88 $\pm$ 0.02 & -7.65 & -8.53 & Kasliwal (10136)\\
56922.78 & 0.065 $\pm$ 0.001 & 0.090 $\pm$ 0.001 & 16.59 $\pm$ 0.02 & 15.76 $\pm$ 0.01 & 0.83 $\pm$ 0.02 & -7.71 & -8.54 & Kasliwal (10136)\\
56951.33 & 0.063 $\pm$ 0.001 & 0.081 $\pm$ 0.001 & 16.62 $\pm$ 0.02 & 15.86 $\pm$ 0.01 & 0.76 $\pm$ 0.02 & -7.68 & -8.44 & Kasliwal (10136)\\
57121.20 & 0.062 $\pm$ 0.001 & 0.082 $\pm$ 0.001 & 16.63 $\pm$ 0.02 & 15.85 $\pm$ 0.02 & 0.78 $\pm$ 0.02 & -7.67 & -8.45 & Kasliwal (11063)\\
57300.46 & 0.064 $\pm$ 0.001 & 0.084 $\pm$ 0.001 & 16.61 $\pm$ 0.02 & 15.82 $\pm$ 0.01 & 0.79 $\pm$ 0.02 & -7.69 & -8.48 & Kasliwal (11063)\\
57305.62 & 0.063 $\pm$ 0.001 & 0.094 $\pm$ 0.001 & 16.62 $\pm$ 0.02 & 15.71 $\pm$ 0.01 & 0.91 $\pm$ 0.02 & -7.68 & -8.59 & Kasliwal (11063)\\
57320.45 & 0.063 $\pm$ 0.001 & 0.083 $\pm$ 0.001 & 16.63 $\pm$ 0.02 & 15.84 $\pm$ 0.01 & 0.79 $\pm$ 0.02 & -7.67 & -8.46 & Kasliwal (11063)\\
57462.10 & 0.064 $\pm$ 0.001 & 0.083 $\pm$ 0.001 & 16.60 $\pm$ 0.02 & 15.83 $\pm$ 0.01 & 0.77 $\pm$ 0.02 & -7.70 & -8.47 & Kasliwal (11063)\\
57469.00 & 0.062 $\pm$ 0.001 & 0.081 $\pm$ 0.001 & 16.63 $\pm$ 0.02 & 15.87 $\pm$ 0.01 & 0.76 $\pm$ 0.02 & -7.67 & -8.43 & Kasliwal (11063)\\
57488.83 & 0.064 $\pm$ 0.001 & 0.082 $\pm$ 0.001 & 16.61 $\pm$ 0.02 & 15.86 $\pm$ 0.01 & 0.75 $\pm$ 0.02 & -7.69 & -8.44 & Kasliwal (11063)\\
57673.33 & 0.062 $\pm$ 0.001 & 0.081 $\pm$ 0.001 & 16.64 $\pm$ 0.02 & 15.86 $\pm$ 0.01 & 0.78 $\pm$ 0.02 & -7.66 & -8.44 & Kasliwal (13053)\\
57852.92 & 0.061 $\pm$ 0.001 & 0.079 $\pm$ 0.001 & 16.67 $\pm$ 0.02 & 15.90 $\pm$ 0.01 & 0.77 $\pm$ 0.02 & -7.63 & -8.40 & Kasliwal (13053)\\
58045.60 & 0.062 $\pm$ 0.001 & 0.081 $\pm$ 0.001 & 16.63 $\pm$ 0.02 & 15.86 $\pm$ 0.01 & 0.77 $\pm$ 0.02 & -7.67 & -8.44 & Kasliwal (13053)\\
58242.81 & 0.091 $\pm$ 0.001 & 0.150 $\pm$ 0.001 & 16.22 $\pm$ 0.02 & 15.20 $\pm$ 0.01 & 1.02 $\pm$ 0.01 & -8.08 & -9.10 & Kasliwal (13053)\\
58443.65 & 0.104 $\pm$ 0.001 & 0.188 $\pm$ 0.001 & 16.08 $\pm$ 0.01 & 14.95 $\pm$ 0.01 & 1.13 $\pm$ 0.01 & -8.22 & -9.35 & Kasliwal (14089)\\
58581.44 & 0.074 $\pm$ 0.001 & 0.111 $\pm$ 0.001 & 16.45 $\pm$ 0.02 & 15.52 $\pm$ 0.01 & 0.93 $\pm$ 0.02 & -7.85 & -8.78 & Kasliwal (14089)\\
58607.97 & 0.074 $\pm$ 0.001 & 0.109 $\pm$ 0.001 & 16.44 $\pm$ 0.02 & 15.54 $\pm$ 0.01 & 0.90 $\pm$ 0.02 & -7.86 & -8.76 & Kasliwal (14089)\\
58795.85 & 0.067 $\pm$ 0.001 & 0.098 $\pm$ 0.001 & 16.56 $\pm$ 0.02 & 15.66 $\pm$ 0.01 & 0.90 $\pm$ 0.02 & -7.74 & -8.64 & Kasliwal (14089)\\
\enddata
\tablecomments{\spitzer/IRAC 3.6 and 4.5 $\mu$m photometry (in Jy and magnitudes), $\scol$ color, and absolute magnitudes with 1$\sigma$ uncertainties of the six dust-forming WC candidates SPIRITS~16ln, 19q, 16df, 14apu, 18hb, and 14bqe. The PI names and program ID numbers of the \spitzer/IRAC programs associated with the observations are also provided. All magnitudes are given in the Vega system.}
\label{tab:LCTabFull}
\end{deluxetable*}
\clearpage



\end{document}